\documentclass[%
 amsmath,amssymb,
 aps,
pra,
floatfix,
]{revtex4-2}

\usepackage{graphicx}
\usepackage{dcolumn}
\usepackage{bm}

\makeatletter
\renewcommand\@makecaption[2]{%
  \par
  \vskip\abovecaptionskip
  \begingroup
   \small\rmfamily
    \begingroup
     \samepage
     \flushing
     \let\footnote\@footnotemark@gobble
     \@make@capt@title{#1}{#2}\par
    \endgroup
  \endgroup
  \vskip\belowcaptionskip
}
\makeatother

\begin{document}
\setlength\textfloatsep{6pt}


\title{Leggett--Garg inequalities with deformed Pegg--Barnett phase observables}

\author{Hiroo Azuma$^{1}$}
\email{hiroo.azuma@m3.dion.ne.jp, zuma@nii.ac.jp}
\author{William J. Munro$^{1,2}$} 
\author{Kae Nemoto$^{1,3}$} 
\affiliation{$^1$Global Research Center for Quantum Information Science,
National Institute of Informatics, 2-1-2 Hitotsubashi, Chiyoda-ku, Tokyo 101-8430, Japan}
\affiliation{$^2$NTT Basic Research Laboratories and Research Center for Theoretical Quantum Physics, 3-1 Morinosato-Wakamiya, Atsugi, Kanagawa 243-0198, Japan}
\affiliation{$^3$Okinawa Institute of Science and Technology Graduate University, Onna-son, Okinawa 904-0495, Japan}




\date{\today}

\begin{abstract}
We investigate the Leggett--Garg inequalities (LGIs) for a boson system
whose observables are given by deforming the Pegg--Barnett phase operators.
We consider two observables and show that the quantum Fourier transform is useful in the realization of the required measurements.
Deriving explicit forms for the LGIs using the coherent state $|\alpha\rangle$ as the initial state,
we explore the regimes where they are violated when the time difference between observations of the phase operators is varied.
We show that the system remains nonclassical in the large amplitude limit
without dissipation, however with dissipation, our violation diminishes rapidly.
\end{abstract}

\maketitle


\section{\label{section-introduction}Introduction}
The Leggett--Garg inequality (LGI) is the temporal analog of Bell's inequality
that shows the differences between quantum mechanics and macroscopic local realism \cite{Leggett1985,Emary2014}.
The LGI was proposed to determine how well the following two assumptions work:
(1) macroscopic realism and (2) noninvasive measurability.
Here macroscopic realism refers to the results of observation that are determined by hidden variables that are qualities of the observed system.
The hidden variables are introduced for describing outcomes of any measurement for the system by a deterministic theory
instead of quantum mechanics that is nondeterministic \cite{Einstein1935}.
Next noninvasive measurability implies that the result of a measurement at a time $t_{2}$ is independent of the result of a previous measurement
at time $t_{1}$($<t_{2}$).
This means we should be able to perform measurements without disturbing the system.
By contrast, quantum mechanics requires an interaction between the system
and the observer meaning the system cannot avoid suffering from effects caused by that observer.
This means we can regard the noninvasive measurability as a characteristic of classical mechanics.
These two assumptions (macroscopic realism and noninvasive measurability) must hold for measurements of classical systems at the macroscopic level.

The above two assumptions for the LGI evoked the following controversies.
Maroney and Timpson \cite{Maroney2014} considered the conflict between quantum theory and macroscopic realism
and pointed out that violation of the LGI might not imply the falsity of macroscopic realism at all.
To test macroscopic realism,
a rigorous experiment of an improved LGI was demonstrated using a superconducting flux qubit \cite{Knee2016}.
Next \cite{Moreira2015} insisted that macroscopic realism was a model-dependent notion
and constructed a model where invasiveness was controlled by a specific parameter.
Then \cite{Clemente2016} indicated that Fine's proof that Bell's inequalities were necessary and sufficient for the existence of a local realistic model
did not hold if we required macroscopic realism \cite{Fine1982}.
Later \cite{Wang2017} demonstrated an experiment that detected the violation of the LGI and excluded macroscopic realism
where all superpositions were statistical mixtures in an appropriate basis.

Despite these controversies,
considerations about the two assumptions mean that we can expect that a violation of the LGI reveals information about quantum temporal correlations.
Further because the LGI is obtained from correlations of the system observed at different times,
collapses of the wave function of the system are essential.
In other words, the collapse of the wave packet contradicts the noninvasive measurability,
so that the LGI must be performed using quantum nondemolition (QND) measurements.
In general, QND measurements collapse the wave function unless it is an eigenstate of the observables.
In our situation, the system's state is not a state meaning our LGI differs from Bell's inequality in terms of this crucial point \cite{Leggett1985,Emary2014}.

We know that in order to determine an explicit form for our LGI, we need to choose our observable.
Which observable we choose is a very difficult problem
from a practical point of view.
Because the LGI requires QND measurements,
a simple quantum circuit for measuring the observable is preferable.
So far, the Pauli operators for a qubit system and the displaced parity operators for a boson system have been used
as the observables \cite{Friedenberger2017,Azuma2021}.
The LGI for a damped qubit system with Pauli operators as the observables was also explored \cite{Friedenberger2017}.
Further, the LGI of a boson system weakly coupled to a zero-temperature environment
with displacement parity operator observables was shown for a coherent initial state to produce a larger violation than the cat state \cite{Azuma2021}.

Next there have been several experimental demonstrations showing an LGI violation
\cite{Palacios-Laloy2010,Knee2012,Xu2011}.
In particular \cite{Xu2011}, the violation of the LGI for a single photon was detected using linear optical QND measurements \cite{Pryde2004}.
QND measurements are however difficult to efficiently implement.
Goggin {\it et al}. experimentally demonstrated the viiolation of a LGI using the weak measurement of photons
as a noninvasive measurement
instead of the QND measurement \cite{Goggin2011}.
A relationship between the weak values and the LGI was explored \cite{Williams2008}.
Finally LGI experiments were undertaken via the modified ideal negative result measurement scheme
on a three-level system in liquid-state nuclear magnetic resonance \cite{Katiyar2017}.

The systems mentioned above have focused on discrete variable encodings
as it had been difficult to find an appropriate Hermitian phase operator for the infinite-dimensional number state Hilbert space \cite{Dirac1927,Susskind1964}.
That difficulty arises from the fact that eigenvalues of the number operator are positive or equal to zero and the existence of the vacuum state is significant.
Pegg and Barnett steered clear of this trouble by considering the ring-shaped finite-dimensional Hilbert space
and set an upper bound of the energy \cite{Pegg1988,Pegg1989,Barnett1989}.
Experimental exploration of the Pegg--Barnett phase operator has been investigated
\cite{Noh1991,Noh1992a,Noh1992b,Beck1993,Smithey1993}
with it being measured using photon counting in \cite{Noh1991,Noh1992a,Noh1992b}.
The uncertainty relation for the phase and photon number of the electromagnetic field in a coherent state
of a small average photon number was experimentally estimated with optical homodyne tomography \cite{Beck1993,Smithey1993}.
However, we cannot apply these methods to QND measurements.

In this work, we explore the LGIs for a boson system weakly coupled to a zero-temperature environment.
In this scenario, we can exactly solve the master equation of the boson system with a closed-form expression
\cite{Azuma2021,Walls1994,Barnett1997}.
We consider two deformed Pegg--Barnett operators as observables of the LGI.
We show that the quantum Fourier transform (QFT) simplifies the implementation of QND measurements considerably.
Further we examine fluctuations of observation of the deformed Pegg--Barnett operator caused
by imperfect gates in a quantum circuit of the QFT.

Let us begin by explaining these observables and the time evolution of the LGI in more detail.
To define the Pegg--Barnett phase operator in a finite-dimensional Hilbert space,
an orthonormal basis is prepared.
We construct two pairs of projectors from this orthonormal basis.
We build two observables from these projectors
(labeled $\hat{O}^{(\mbox{\scriptsize parity})}_{s}$ and $\hat{O}^{(+,-)}_{s}$).
Assuming we begin with an initial coherent state and that the boson system evolves under weak interaction with a zero-temperature environment,
we can derive explicit forms of the LGIs of these two observables.
To estimate the violation of the LGI, we need to measure the observable at different times.
We regard the violation of the LGI as a function of the time between the two measurements in the LGI protocol.
If we assume that there is no dissipation in the system,
the function is periodic.
Now to evaluate the quantumness of the system,
we compute a dimensionless quantity which is the proportion of the LGI violation in the total period.
Next it is important to mention that when dissipation is introduced,
the function is not periodic.
However, even in this case, we can still compute this quantity.
We show that without dissipation,
this quantity remains at a nonzero value and the violation of the LGI does not vanish in a specific limit.
By contrast,
if the system has dissipation,
the violation of the LGI disappears and the dimensionless quantity converges to zero as the degree of the dissipation becomes larger.
This is one of the major results of our work.

Now to the structure of this paper.
It is organized as follows.
In Sec.~\ref{section-backgrounds},
we provide the theoretical backgrounds for our model, the LGI, and the Pegg--Barnett phase operator.
Next in Sec.~\ref{section-observables-Pegg-Barnett},
we define the observables for the LGIs by deforming the Pegg--Barnett phase operator while in Sec.~\ref{section-implementation-observables},
we discuss the implementation of measurements of the deformed Pegg--Barnett operators using the QFT.
In Sec.~\ref{section-dissipation-QFT},
we evaluate dissipation of measurements of an imperfect quantum circuit of the QFT while in Sec.~\ref{section-derivation-LGI-parity-numerical-calculations},
we derive an explicit form of the LGI with $\hat{O}^{(\mbox{\scriptsize parity})}_{s}$ and provide its numerical evidence.
In Sec.~\ref{section-derivation-LGI-plusminus-numerical-calculations},
we derive an explicit form of the LGI with $\hat{O}^{(+,-)}_{s}$ and provide its numerical evidence.
Finally in Sec.~\ref{section-discussions},
we provide a concluding discussion and summary of our results.

\section{\label{section-backgrounds}Theoretical background}
Before we proceed to our main results let us provide some preliminary materials beginning with our model.
Consider the following master equation of a boson system coupled weakly to a zero-temperature environment:
\begin{equation}
\dot{\rho}(t)=-i\omega[a^{\dagger}a,\rho(t)]
+
\Gamma
\Bigl(
2a\rho(t)a^{\dagger}-a^{\dagger}a\rho(t)-\rho(t)a^{\dagger}a
\Bigr),
\label{master-equation-0}
\end{equation}
where $a$ and $a^{\dagger}$ are the annihilation and creation operators of the boson system,
$\Gamma$ is the decay rate,
and $\omega$ is the angular frequency of the boson system.
Here, we draw attention to the fact that we do not work in the interaction picture where we could eliminate the Hamiltonian term.
Instead we will work in the Schr{\"{o}}dinger picture.
The decay rate $\Gamma$ and the angular frequency of the boson system $\omega$ have the dimension of reciprocal time.
In Appendix~\ref{Appendix-A}, we explicitly show the solution of such a master equation.

Next, the LGI is defined as follows.
First, we consider an operator $\hat{O}$ whose eigenvalues are given by $\pm 1$.
Second, we introduce three
equally spaced measurement times,
$t_{1}=0$, $t_{2}=\tau$, and $t_{3}=2\tau$ with $\tau>0$.
Third, we describe an observed value of $\hat{O}$ at $t_{i}$ as $O_{i}$.
Fourth, we write the probability that we obtain observed values $O_{i}$ and
$O_{j}$ at times $t_{i}$ and $t_{j}(>t_{i})$ respectively as $P_{ij}(O_{i},O_{j})$.
Fifth, we define the correlation function $C_{ij}$ as
\begin{equation}
C_{ij}
=
\sum_{O_{i},O_{j}\in\{-1,+1\}}O_{i}O_{j}P_{ij}(O_{i},O_{j}).
\end{equation}
Finally, we can write the LGI as follows:
\begin{equation}
-3
\leq
K_{3}
=
C_{12}+C_{23}-C_{13}
\leq
1.
\label{K3-formula-0}
\end{equation}

Now to define the Pegg--Barnett phase operator
\cite{Pegg1988,Pegg1989,Barnett1989}, we consider the following state vectors:
\begin{equation}
|\phi_{m}\rangle_{s}
=
(s+1)^{-1/2}\sum_{n=0}^{s}\exp[in\phi_{m}]|n\rangle,
\label{definition-Pegg-Barnett-basis-vector-0}
\end{equation}
where
\begin{equation}
\phi_{m}
=
\phi_{0}+\frac{2\pi}{s+1}m
\quad
\mbox{for $m=0,1,2,...,s$},
\end{equation}
and $\{|n\rangle:n=0, ..., s\}$ are Fock number states.
The vectors
$\{|\phi_{m}\rangle_{s}:m=0,1,2,...,s\}$ form a complete orthonormal system in the $(s+1)$-dimensional Hilbert space.
Then, the Pegg--Barnett phase operator can be defined as \cite{Pegg1988,Pegg1989,Barnett1989}:
\begin{equation}
\hat{\phi}
=
\sum_{m=0}^{s}\phi_{m}|\phi_{m}\rangle_{s}{}_{s}\langle\phi_{m}|.
\label{Pegg-Barnett-phase-operator-0}
\end{equation}

\section{\label{section-observables-Pegg-Barnett}Observables from the deformed Pegg--Barnett phase operator}
We can use the basis $\{|\phi_{m}\rangle_{s}\}$ to define two observables,
$\hat{O}^{(\mbox{\scriptsize parity})}_{s}$ and $\hat{O}^{(+,-)}_{s}$ which are given by
\begin{equation}
\hat{O}^{(\mbox{\scriptsize parity})}_{s}
=
\Pi^{(\mbox{\scriptsize even})}_{s}-\Pi^{(\mbox{\scriptsize odd})}_{s},
\label{orthogonal-measurement-operator-parity}
\end{equation}
\begin{equation}
\hat{O}^{(+,-)}_{s}
=
\Pi^{(+)}_{s}-\Pi^{(-)}_{s},
\label{orthogonal-measurement-operator-0}
\end{equation}
with
\begin{equation}
\Pi^{(\mbox{\scriptsize even})}_{s}
=
\sum_{m=0}^{N}|\phi_{2m}\rangle_{s}{}_{s}\langle\phi_{2m}|,
\quad
\Pi^{(\mbox{\scriptsize odd})}_{s}
=
\sum_{m=0}^{N}|\phi_{2m+1}\rangle_{s}{}_{s}\langle\phi_{2m+1}|, 
\end{equation}
\begin{equation}
\Pi^{(+)}_{s}
=
\sum_{m=0}^{N}|\phi_{m}\rangle_{s}{}_{s}\langle\phi_{m}|,
\quad
\Pi^{(-)}_{s}
=
\sum_{m=N+1}^{2N+1}|\phi_{m}\rangle_{s}{}_{s}\langle\phi_{m}|.
\end{equation}
Here $s=2N+1$ with $N=0,1,2,...$.
Remembering that $\{|\phi_{m}\rangle_{s}\}$ form a complete orthonormal basis,
the eigenvalues of $\hat{O}^{(\mbox{\scriptsize parity})}_{s}$ and $\hat{O}^{(+,-)}_{s}$ must be either $\pm 1$.
Now in our work here we will let $s$ be large but not infinite.
Because the right-hand side of Eq.~(\ref{definition-Pegg-Barnett-basis-vector-0}) is divided by $\sqrt{s+1}$,
some might worry whether taking the large $s$ limit is allowed.
However, a relationship
$\mbox{Tr}[\Pi^{(\mbox{\scriptsize even})}_{s}]
=
\mbox{Tr}[\Pi^{(\mbox{\scriptsize odd})}_{s}]
=
\mbox{Tr}[\Pi^{(+)}_{s}]
=
\mbox{Tr}[\Pi^{(-)}_{s}]
=
1/2$
holds rigorously,
so that the factor $(s+1)^{-1/2}$ does not cause issues.

\section{\label{section-implementation-observables}Measurements with the
observables $\hat{O}^{(\mbox{\scriptsize parity})}_{s}$ and $\hat{O}^{(+,-)}_{s}$}
If we carry out measurements with the projection operators constructed from $\{|\phi_{m}\rangle_{s}\}$,
we can perform measurements with $\hat{O}^{(\mbox{\scriptsize parity})}_{s}$ and $\hat{O}^{(+,-)}_{s}$.
Although this method is not always optimal for determining $\hat{O}^{(\mbox{\scriptsize parity})}_{s}$ and $\hat{O}^{(+,-)}_{s}$,
we will proceed in that direction as it gives us a clear perspective.
From Eq.~(\ref{definition-Pegg-Barnett-basis-vector-0}),
we can regard the vector $|\phi_{m}\rangle_{s}$ as a superposition of the boson system's number states $\{|n\rangle:n=0,1,...,s\}$.
Here, we set
$s=d-1$ with $d=2^{L}$.
Remembering that $s=2N+1$, we have $N=(d/2)-1$.
If we put $L=1,2,3,4,...$,
we need to set $s=1,3,7,15,...$.
This in turn obviously means the dimension of the Hilbert space on which $\hat{O}^{(\mbox{\scriptsize parity})}_{s}$ and $\hat{O}^{(+,-)}_{s}$
are defined is restricted by the number of qubits for the QFT.

We can now consider the following unitary transformation that maps a number state of the boson system to a state of $L$ qubits:
\begin{equation}
U:
|x\rangle
\to
|x_{1}\rangle
|x_{2}\rangle
...
|x_{L}\rangle
\quad
\mbox{for $x=0,1,...,d-1$},
\label{boson-qubit-transformation-0}
\end{equation}
with
$x_{i}\in\{0,1\}$ for $i=1, 2, ..., L$ and $x=\sum_{i=1}^{L}x_{i}2^{L-i}$.

Realization of this transformation $U$ is challenging and beyond the scope of the article.
By postulating the existence of $U$,
we can realize measurements of $\hat{O}^{(\mbox{\scriptsize parity})}_{s}$ and $\hat{O}^{(+,-)}_{s}$ however as follows.
First, we consider the QFT of $L$ qubits \cite{Coppersmith1994,Ekert1996},
\begin{equation}
\mbox{QFT}:
|x\rangle
\to
\frac{1}{\sqrt{d}}
\sum_{y=0}^{d-1}
\exp[i(2\pi/d)xy]|y\rangle,
\label{definition-QFT-0}
\end{equation}
for $x=0, 1, ..., d-1$.
Second, we apply the QFT to $|\phi_{m}\rangle_{s}$ as follows:
\begin{equation}
\mbox{QFT}:
|\phi_{m}\rangle_{s}
\to
\frac{1}{d}
\sum_{y=0}^{d-1}
\frac{1-\exp(izd)}{1-\exp(iz)}
|y\rangle,
\label{phi-m-QFT-0}
\end{equation}
\begin{equation}
z
=
\phi_{0}
+
\frac{2\pi}{d}(y+m).
\label{z-definition}
\end{equation}
We draw attention to the fact that the $\sqrt{s+1}$ part
in Eq.~(\ref{definition-Pegg-Barnett-basis-vector-0}) has been integrated into a factor $1/d$
in the right-hand side of Eq.~(\ref{phi-m-QFT-0})
because $s+1=d$.
The right-hand side of Eq.~(\ref{phi-m-QFT-0}) has sharp peaks around $z=0$.
Thus, if we observe the right-hand side of Eq.~(\ref{phi-m-QFT-0}),
we obtain $|y\rangle$ with a probability of almost unity where $y$ is the nearest integer of $\{-m-[d/(2\pi)]\phi_{0}\}$.
Hence, if we put $\phi_{0}=0$,
$y=m$ holds.
Therefore, the QFT maps each $|\phi_{m}\rangle_{s}$ into a computational basis vector $|m\rangle$ for $m=0,1,...,d-1$.
This implies that we can identify $|\phi_{m}\rangle_{s}$ with ease.

Now to observe the observable $\hat{O}^{(\mbox{\scriptsize parity})}_{s}$,
we need to measure only one qubit $|y_{1}\rangle$ among the $L$ qubits.
This is because we must only distinguish whether $y=\sum_{i=1}^{L}y_{i}2^{L-i}$ is odd or even.
In contrast, to observe the observable $\hat{O}^{(+,-)}_{s}$,
we have to measure only one qubit $|y_{L}\rangle$ among the $L$ qubits.
This is because we must only distinguish whether or not $y$ is smaller than $N+1=d/2$.
The fact that we can observe $\hat{O}^{(\mbox{\scriptsize parity})}_{s}$ and
$\hat{O}^{(+,-)}_{s}$
by detecting only the one qubit gives us a great advantage in carrying out the QND measurements.
Next Fig.~\ref{figure-01ab}~(a) shows how to perform the QND measurement of $\hat{O}^{(\mbox{\scriptsize parity})}_{s}$.
First, we apply the QFT to a state $|\psi\rangle=\sum_{m}c_{m}|\phi_{m}\rangle_{s}$
followed by measuring $|y_{1}\rangle$ in the basis $\{|0\rangle,|1\rangle\}$.
Then we insert $|0\rangle$ or $|1\rangle$ into the first qubit according to the result of the measurement leaving $(L-1)$ qubits untouched,
and feed all of the qubits into the inverse QFT.
Finally, we obtain a wave packet that is generated by a collapse,
$\Pi^{(\chi)}_{s}|\psi\rangle/\sqrt{\langle\psi|\Pi^{(\chi)}_{s}|\psi\rangle}$
for
$\chi\in\{\mbox{even},\mbox{odd}\}$.
We can perform the QND measurement with $\hat{O}^{(+,-)}_{s}$ similarly.

\begin{figure}[ht]
\centering
\begin{minipage}[b]{0.45\linewidth}
\begin{center}
\includegraphics[keepaspectratio, scale=0.8]{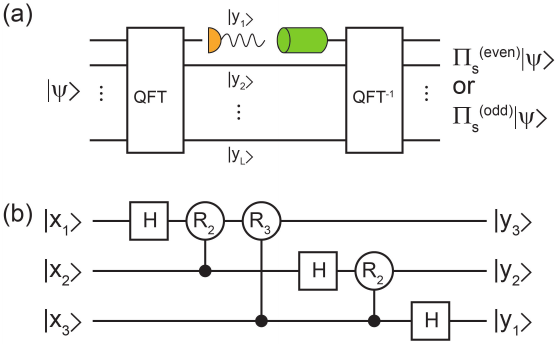}
\end{center}
\caption{
(a) Schematic of the QND measurement with $\hat{O}^{(\mbox{\scriptsize parity})}_{s}$.
(b) Quantum circuit of three-qubit QFT.
The Hadamard gate and the controlled phase shift gate are represented by $H$ and $R_{k}$, respectively.
Here, $R_{k}$ rotates the phase by $2\pi/2^{k}$.
}
\label{figure-01ab}
\vspace{10\baselineskip}
\end{minipage}
\hspace{0.04\columnwidth}
\begin{minipage}[b]{0.45\linewidth}
\begin{center}
\includegraphics[keepaspectratio, scale=0.58]{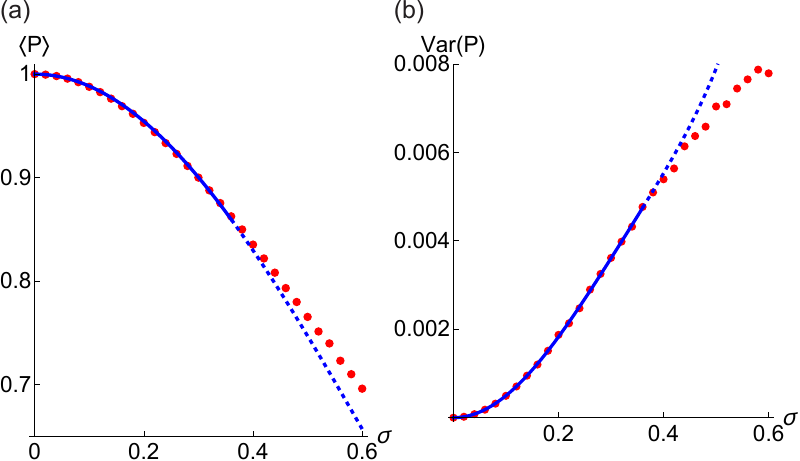}
\end{center}
\caption{Monte Carlo simulations of $\langle P\rangle$ and $\mbox{Var}(P)$.
In (a) we plot $\langle P\rangle$ as a function of $\sigma$
where the red points represent simulation results.
Because an error of each result is within $\pm 2.6\times 10^{-3}$,
we do not put error bars onto our simulation points.
Our approximate solution given by Eq.~(\ref{average-fitting-0}) is shown with the solid blue curve for $0\leq\sigma\leq 0.36$
and the dashed blue curve for $0.36\leq\sigma\leq 0.6$.
(b) Plot of the $\mbox{Var}(P)$ as a function of $\sigma$.
Because an error of each result is within $\pm 2.5\times 10^{-4}$,
we do not put error bars onto our simulation points.
Our approximate solution from Eq.~(\ref{variance-fitting-0}) is depicted with the solid blue curve for $0\leq\sigma\leq 0.36$ and the dashed blue curve for $0.36\leq\sigma\leq 0.6$.
For both the plots shown in (a) and (b),
the fitted curves of Eqs.~(\ref{average-fitting-0}) and (\ref{variance-fitting-0}) are effective for small $\sigma$,
that is,
$0\leq\sigma\leq 0.36$.
This is because these approximate curves are reliable under perturbation theory.
As explained before
the $\mbox{noisy C-}R_{k}$ is useful for $\Delta_{pq}<2\pi/d=2\pi/2^{L}$.
Because we put $L=5$, the fitting parameters $c_{0}$, $c_{1}$, $c_{2}$, and $c_{3}$ are trustworthy for $\sigma<2\pi/d=0.196$.
}
\label{figure-02ab}
\end{minipage}
\end{figure}

As mentioned before, one of the research tasks from now on is finding out how to implement the unitary transformation $U$
given by Eq.~(\ref{boson-qubit-transformation-0}).
We cannot omit it before performing the QFT.
We do not examine a method for the realization of this boson--qubit transformation $U$ in detail because it is very difficult
and lies outside the scope of our study.

\section{\label{section-dissipation-QFT}Errors and imperfections in the measurement of
$\hat{O}^{(+,-)}_{s}$}
The quantum circuit that carries out the QFT is composed of the Hadamard gates and controlled phase shift gates and so in
Fig.~\ref{figure-01ab}~(b) we show an illustrative three-qubit QFT.
In general, for most of the suggested realizations, for example, trapped ion quantum computers,
the single-qubit operations are faster than the conditional gates for two-qubit operations which has implications in terms of errors and imperfections.
We can assume that the conditional phase shift gates are more vulnerable to decoherence than the Hadamard gates
\cite{Barenco1996} and so it is convenient for illustrative purposes to introduce random noise only to the conditional phase shift gates.
For example, letting $|x\rangle_{1}$ and $|y\rangle_{2}$ represent the first and second qubits and regarding them as the control and target qubits respectively,
we can describe an imperfect conditional phase shift gate as
\begin{equation}
\mbox{noisy C-}R_{k}:
|x\rangle_{1}|y\rangle_{2}
\to
\left\{
\begin{array}{ll}
|x\rangle_{1}|y\rangle_{2} & \mbox{for $(x,y)\neq (1,1)$} \\
\exp[(i2\pi/2^{k})+i\Delta_{12}]|x\rangle_{1}|y\rangle_{2} & \mbox{for $(x,y)=(1,1)$} \\
\end{array}
\right.,
\label{noisy-C-Rk-0}
\end{equation}
where $\Delta_{12}$ is a Gaussian random variable with mean zero and variance $\sigma^{2}$ whose probability is given by
\begin{equation}
P(\Delta_{12})
=
\frac{1}{\sqrt{2\pi}\sigma}
\exp
\Bigl[
-\frac{\Delta_{12}^{2}}{2\sigma^{2}}
\Bigr].
\label{Gaussian-probability-formula-0}
\end{equation}
It is then straightforward to obtain the noisy QFT for $L$ qubits as
\begin{equation}
\mbox{noisy QFT}:
|x\rangle
\to
\frac{1}{\sqrt{d}}
\sum_{y=0}^{d-1}
\exp
\Biggl[
i(2\pi/d)xy
+i
\sum_{\substack{{p\in\{2,...,L\},}\\{q\in\{1,...L-1\},}\\{p>q}}}
\Delta_{pq}x_{p}y_{L-q+1}
\Biggr]
|y\rangle.
\label{definition-noisy-QFT-0}
\end{equation}
To estimate the variance of the probability of the wave packets obtained by measurements of $\hat{O}^{(+,-)}_{s}$
with the imperfect QFT,
we need to define our initial state.
We use the following $L$-qubit state that is similar to the coherent state:
\begin{equation}
|\alpha\rangle_{s}
=
C_{s}(\alpha)
\sum_{x=0}^{d-1}
\frac{\alpha^{x}}{\sqrt{x!}}|x\rangle,
\label{definition-alpha-s-coherent}
\end{equation}
\begin{equation}
C_{s}(\alpha)
=
\Biggl(
\sum_{x=0}^{d-1}
\frac{|\alpha|^{2x}}{x!}
\Biggr)^{-1/2}.
\end{equation}
We draw attention to the fact that the normalization factor $C_{s}(\alpha)$ depends on $s=d-1$
because $|\alpha\rangle_{s}$ lies on $(s+1)$-dimensional Hilbert space.

Now let us consider the time evolution of $|\alpha\rangle_{s}$ as follows.
First,
the noisy QFT transforms $|\alpha\rangle_{s}$ as
\begin{equation}
|\psi\rangle_{s}=(\mbox{noisy QFT})|\alpha\rangle_{s}.
\end{equation}
Noting that we want to observe $\hat{O}^{(+,-)}_{s}$,
we should therefore measure $|y_{L}\rangle$.
If we detect $y_{L}=0$ corresponding to $O^{(+,-)}_{s}=1$,
the wave function collapses from $|\psi\rangle_{s}$ to the non-normalized state
\begin{equation}
|\psi_{+}\rangle_{s}=|y_{L}=0\rangle\langle y_{L}=0|\psi\rangle_{s},
\end{equation}
where
$|y_{L}=0\rangle\langle y_{L}=0|$ is a one-qubit projector.
In contrast,
if we detect $y_{L}=1$ corresponding to $O^{(+,-)}_{s}=-1$,
the wave function collapses from $|\psi\rangle_{s}$ to the non-normalized state
\begin{equation}
|\psi_{-}\rangle_{s}=|y_{L}=1\rangle\langle y_{L}=1|\psi\rangle_{s}.
\end{equation}
Explicit forms of $|\psi\rangle_{s}$ and $|\psi_{\pm}\rangle_{s}$ are given in Appendix~\ref{Appendix-B}.

Although we are examining decoherence,
we describe states that suffer from noises of imperfect gates as pure states,
$|\psi_{+}\rangle_{s}$ and $|\psi_{-}\rangle_{s}$.
This is because the error of the noisy conditional phase shift gate is introduced by the unitary operation.
However, it is all right because the states will become mixed by carrying out many runs for the Monte Carlo simulations.
To accomplish the QND measurements,
we need to apply $(\mbox{noisy QFT})^{-1}$ to $|\psi_{\pm}\rangle_{s}$ as
\begin{equation}
|\psi'_{\pm}\rangle_{s}=(\mbox{noisy QFT})^{-1}|\psi_{\pm}\rangle_{s}.
\label{psi-dash-pm-0}
\end{equation}
Explicit forms of $|\psi'_{\pm}\rangle_{s}$ are given in Appendix~\ref{Appendix-B}.

Now to evaluate the noisy operation on our measurement of $\hat{O}^{(+,-)}_{s}$,
we consider the following probability:
\begin{equation}
P(\sigma)
=
|
_{s}\langle \psi'_{+}(\Delta_{pq}=\tilde{\Delta}_{pq}=0)|\psi'_{+}(\Delta_{pq},\tilde{\Delta}_{pq})\rangle_{s}
+
_{s}\langle \psi'_{-}(\Delta_{pq}=\tilde{\Delta}_{pq}=0)|\psi'_{-}(\Delta_{pq},\tilde{\Delta}_{pq})\rangle_{s}
|^{2}, \nonumber \\
\label{imperfect-gates-P-sigma-0}
\end{equation}
where
$\{\Delta_{pq}\}$ and $\{\tilde{\Delta}_{pq}\}$ are Gaussian variables
generated by the noisy QFT and $(\mbox{noisy QFT})^{-1}$ respectively,
and
the relationships between $\sigma$ and $\{\Delta_{pq},\tilde{\Delta}_{pq}\}$ are given by Eq.~(\ref{Gaussian-probability-formula-0}).
Obviously, $P(\sigma=0)=1$ holds.
We can now define the average and variance of $P(\sigma)$ as $\langle P\rangle$ and
$\mbox{Var}(P)=\langle(P-\langle P\rangle)^{2}\rangle$, respectively.
We obtain $\langle P\rangle$ and $\mbox{Var}(P)$ using Monte Carlo simulations with $2\times 10^{4}$ samples
for the $L=5$ qubit case with $\alpha=1/2$.
In Figs.~\ref{figure-02ab}~(a) and (b),
we plot the average and the variance as functions of $\sigma$, respectively.
Fitting the function $\langle P\rangle=\exp(-c_{0}\sigma^{2})$
chosen due to the rotational error's nature to the red data points shown in Fig.~\ref{figure-02ab} (a) for $0\leq \sigma\leq 0.36$,
we obtain
\begin{equation}
c_{0}
=
-1.168\pm 0.0023,
\label{average-fitting-0}
\end{equation}
as an approximation.
This approximation holds for small $\sigma$.
Looking at Eq.~(\ref{noisy-C-Rk-0}),
we notice that the error of the conditional phase shift gate is crucial when
$\Delta_{pq}\sim 2\pi/d$.
Because of $L=5$ and Eq.~(\ref{Gaussian-probability-formula-0}),
we estimate the typical $\sigma$ at $2\pi/2^{5}=0.196$.

In Fig.~\ref{figure-02ab}~(b), we show our numerical results for $\mbox{Var}(P)$ by the red dots.
Fitting a polynomial function
$\mbox{Var}(P)
=
c_{1}\sigma^{2}
+
c_{2}\sigma^{4}
+
c_{3}\sigma^{6}$
to the data points for $0\leq\sigma\leq 0.36$, we obtain
\begin{eqnarray}
c_{1}
&=&
0.051{\,}22\pm 0.001{\,}127, \nonumber \\
c_{2}
&=&
-0.1499\pm 0.034{\,}59, \nonumber \\
c_{3}
&=&
0.2850\pm 0.2514.
\label{variance-fitting-0}
\end{eqnarray}

We can witness similar effects for the measurements of the operator $\hat{O}^{(\mbox{\scriptsize parity})}_{s}$.

\section{\label{section-derivation-LGI-parity-numerical-calculations}
Derivation of the LGI of the observable
$\hat{O}^{(\mbox{\scriptsize parity})}_{s}$}
Let us now turn our attention to the LGIs and their potential violation.
We start by preparing the coherent state $|\alpha\rangle$ as our initial state of the system at $t_{1}=0$.
Choosing $\hat{O}^{(\mbox{\scriptsize parity})}_{s}$ as our observable and based on the details presented in Appendix~\ref{Appendix-C}
we can obtain the correlation function $C_{12}$ as
\begin{equation}
C_{12}
=
\exp[-|\alpha|^{2}]
\mbox{Re}
\Bigl[
\mbox{Tr}[\Pi^{(\mbox{\scriptsize even})}_{s}(L_{1}(\tau)+L_{2}(\tau))]
-
\mbox{Tr}[\Pi^{(\mbox{\scriptsize odd})}_{s}(L_{1}(\tau)+L_{2}(\tau))]
\Bigl],
\label{C21-parity-formula-0}
\end{equation}
where
\begin{eqnarray}
&&
\mbox{Tr}[\Pi^{(\mbox{\scriptsize even})}_{s}(L_{1}(\tau)+L_{2}(\tau))]
-
\mbox{Tr}[\Pi^{(\mbox{\scriptsize odd})}_{s}(L_{1}(\tau)+L_{2}(\tau))] \nonumber \\
&=&
\sum_{n=0}^{N}
\{
\frac{|\alpha|^{2n}}{n!}
\exp[i\omega(N+1)\tau]
+
\frac{|\alpha|^{2(n+N+1)}}{(n+N+1)!}
\exp[-i\omega(N+1)\tau]
\} \nonumber \\
&&
\times
\exp[-\Gamma \tau(2n+N+1)]
\sum_{l=0}^{n}
\Bigl[
\exp[2\Gamma t]-1
\Bigr]^{l}
\sqrt{
\left(
\begin{array}{c}
n+N+1 \\
l \\
\end{array}
\right)
\left(
\begin{array}{c}
n \\
l \\
\end{array}
\right)
},
\label{Trevenodd-L1L2-tau}
\end{eqnarray}
and $L_{1}(\tau)$ and $L_{2}(\tau)$ are given by Eqs.~(\ref{L1-tau-parity}) and (\ref{L2-tau-parity}), respectively.
In Eq.~(\ref{C21-parity-formula-0}),
the real part of Eq.~(\ref{Trevenodd-L1L2-tau}) remains with the reason given in Appendix~\ref{Appendix-C}.
From Eqs.~(\ref{C21-parity-formula-0}) and (\ref{Trevenodd-L1L2-tau}), we can compute $C_{12}$.
We draw attention to the fact that $C_{12}$ does not depend on $\phi_{0}$.
Instead it is a function of variables $N$, $\tau$, $\alpha$, $\Gamma$, and $\omega$,
that is to say, $C_{12}(N, \tau, \alpha, \Gamma, \omega)$.
We can then obtain $C_{13}$ and $C_{23}$ as
\begin{equation}
C_{13}
=
C_{12}(N, 2\tau, \alpha, \Gamma, \omega),
\end{equation}
\begin{equation}
C_{23}
=
C_{12}(N, \tau, \alpha\exp(-i\Omega\tau), \Gamma, \omega),
\end{equation}
where we used Eq.~(\ref{time-evolution-alpha-beta-0}).
These correlations functions thus enable us to determine $K_{3}$.

\begin{figure}[ht]
\centering
\begin{minipage}[b]{0.45\linewidth}
\begin{center}
\includegraphics[keepaspectratio, scale=0.82]{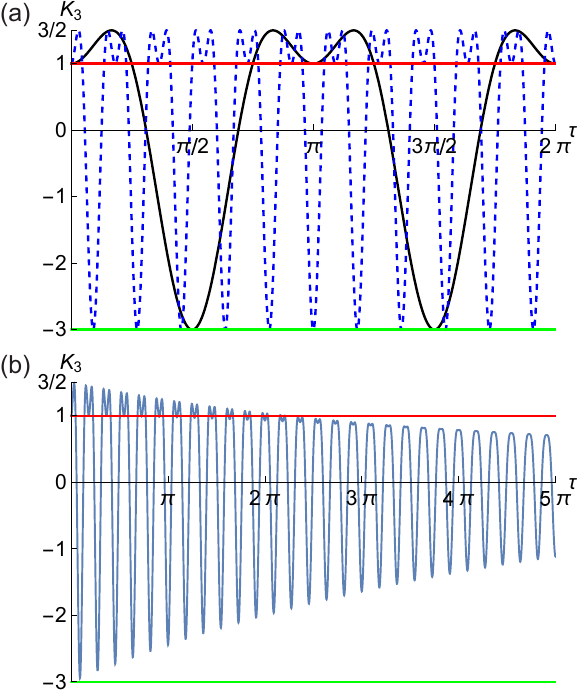}
\end{center}
\caption{Plots of $K_{3}$ as functions of $\tau$.
In (a) we have set $\phi_{0}=0$, $\alpha=1/2$, $\Gamma=0$, and $\omega=1$.
The solid black and dashed blue curves represent $N=1$ and $N=10$, respectively.
A period of the graphs is equal to $2\pi/(N+1)$.
The red and green lines represent $K_{3}=1$ and $K_{3}=-3$, respectively.
These graphs suggest that the period of the curve of $N=1$ is $11/2$ times longer than that of the curve of $N=10$.
In (b) we have set $N=10$, $\phi_{0}=0$, $\alpha=1/2$, $\Gamma=0.005$, and $\omega=1$.}
\label{figure-03ab}
\end{minipage}
\hspace{0.04\columnwidth}
\begin{minipage}[b]{0.45\linewidth}
\begin{center}
\includegraphics[keepaspectratio, scale=0.82]{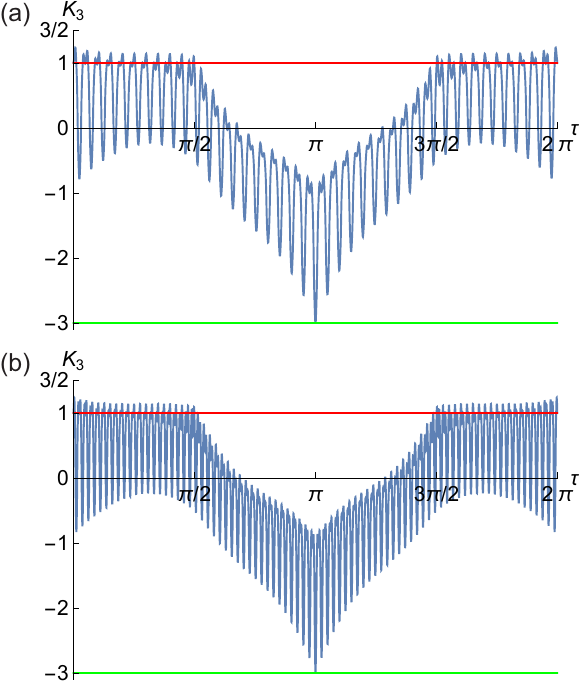}
\end{center}
\caption{Plots of $K_{3}$ as a function of $\tau$
with $\phi_{0}=0$, $\alpha=1/2$, $\Gamma=0$ and $\omega=1$ for (a) $N=20$, (b) $N=40$.}
\label{figure-04ab}
\vspace{5\baselineskip}
\end{minipage}
\end{figure}

It is now useful to numerically explore the behavior of $K_{3}$ using $\hat{O}^{(\mbox{\scriptsize parity})}_{s}$
as our measurement observable with $|\alpha\rangle$ as our initial state.
Figure~\ref{figure-03ab}~(a) shows graphs of $K_{3}$ without dissipation as a function of $\tau$.
The maximum value of $K_{3}$ in Fig.~\ref{figure-03ab}~(a) is equal to $3/2$
with very small numerical errors.
In Fig.~\ref{figure-03ab}~(b), we plot a graph of $K_{3}$ as a function of $\tau$
with nonzero $\Gamma$.
The amplitude of oscillations of the graph decreases gradually as $\tau$ becomes larger
until we reach a stage when the LGI is not violated.

The plots shown in Fig.~\ref{figure-03ab} are very typical for the LGI.
The time evolution of a qubit whose Hamiltonian was proportional to $\sigma_{x}$ exhibited the violation of the LGI
and its graph as a function of $\tau$ was very similar to Fig.~\ref{figure-03ab} (a) \cite{Emary2014}.
In Ref.~\cite{Lambert2011}, the violation of the LGI for an optoelectromechanical system was investigated and its graph as a function of $\tau$
showed modulation caused by the non-energy-conserving terms in interaction.
In this study, the graph resembled Fig.~\ref{figure-03ab} (b), as well.

Next \cite{Kofler2007} showed that the LGI was violated for systems with arbitrary spin lengths.
References~\cite{Kofler2007} and\cite{Kofler2008} pointed out that the time evolution with either decoherence or coarse-grained measurements increased classical properties
and undermined the violation of the LGI.
These considerations were consistent with Fig.~\ref{figure-03ab} (b).
Then \cite{Budroni2014} discussed that the quantum bound of the violation of the LGI depended on the number of levels of the system.
In the limit where the number of levels got closer to infinity,
the violation of the LGI approached the algebraic maximum.
In these references, the LGI for spin systems was investigated.
Figure~\ref{figure-03ab} tells us that the violation of the LGI for the boson system reveals similar properties to multilevel systems.
However, as mentioned before, the realization of the QND measurements for the boson system is very difficult compared to that of the multilevel systems.
This is because we can hardly implement the unitary transformation $U$ defined in Eq.~(\ref{boson-qubit-transformation-0}).
Reference~\cite{Emary2014} introduced an experiment of the LGI with a superconducting transmon qubit \cite{Knee2012}.
In summary, the QND measurements of spin systems have already been viable tasks.
By contrast, counting the number of photons in a QND manner is a difficult task in the future.

\section{\label{section-derivation-LGI-plusminus-numerical-calculations}
Derivation of the LGI of the observable
$\hat{O}^{(+,-)}_{s}$}
Similar to our procedure for the parity observable we can determine what happens for $\hat{O}^{(+,-)}_{s}$ as well
when using the initial coherent state $|\alpha\rangle$.
Choosing $\hat{O}^{(+,-)}_{s}$ as the observable,
based on calculations given in Appendix~\ref{Appendix-D},
we can describe the correlation function $C_{12}$ as
\begin{eqnarray}
C_{12}
&=&
4
\exp[-|\alpha|^{2}]
\mbox{Re}
\Bigl[
\mbox{Tr}[\Pi^{(+)}_{s}L(\tau)]
-
\mbox{Tr}[\Pi^{(-)}_{s}L(\tau)]
\Bigr], \nonumber \\
\label{C21-formula-0}
\end{eqnarray}
where
\begin{eqnarray}
&&
\mbox{Tr}[\Pi^{(+)}_{s}L(\tau)]
-
\mbox{Tr}[\Pi^{(-)}_{s}L(\tau)] \nonumber \\
&=&
4(s+1)^{-2}
\sum_{m=0}^{s}
\sum_{\substack{{m'=0,}\\{m-m'=\mbox{\scriptsize odd}}}}^{s}
\sum_{\substack{{n=0,}\\{n-m'=\mbox{\scriptsize odd}}}}^{s}
\frac{\exp[-i(n-m)\phi_{0}]}{\{\exp[-i(n-m')\pi/(N+1)]-1\}\{\exp[i(m-m')\pi/(N+1)]-1\}} \nonumber \\
&&
\times
\frac{\alpha^{n}\alpha^{*m}}{\sqrt{n!m!}}
\exp[-\Gamma \tau(n+m')-i\omega(n-m')\tau]
\sum_{l=0}^{\mbox{\scriptsize min}(n,m')}
\Bigl[
\exp[2\Gamma \tau]-1
\Bigr]^{l}
\sqrt{
\left(
\begin{array}{c}
n \\
l \\
\end{array}
\right)
\left(
\begin{array}{c}
m' \\
l \\
\end{array}
\right)
},
\label{Trpm-L-tau}
\end{eqnarray}
and $L(\tau)$ is given by Eq.~(\ref{L-tau}).
Similarly to the previous situation we can obtain $C_{13}$ and $C_{23}$ as
\begin{equation}
C_{13}
=
C_{12}(N, 2\tau, \phi_{0}, \alpha, \Gamma, \omega),
\end{equation}
\begin{equation}
C_{23}
=
C_{12}(N, \tau, \phi_{0}, \alpha\exp(-i\Omega\tau), \Gamma, \omega),
\end{equation}
allowing us to determine $K_{3}$.

\begin{figure}[ht]
\centering
\begin{minipage}[b]{0.45\linewidth}
\begin{center}
\includegraphics[keepaspectratio, scale=0.82]{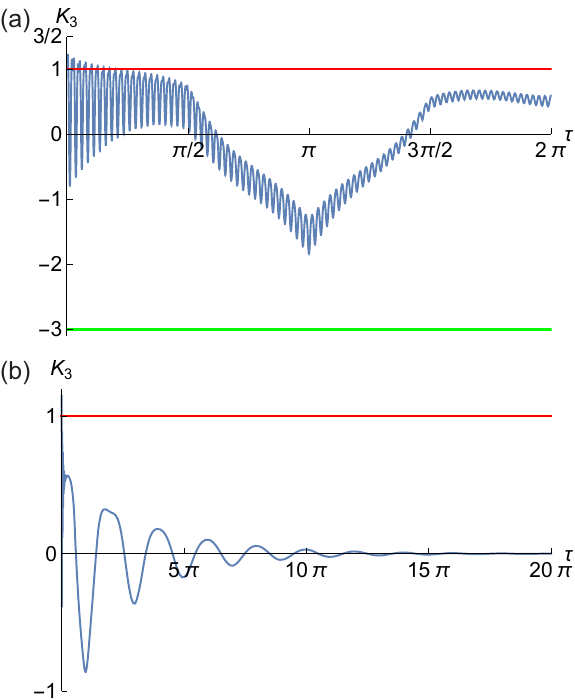}
\end{center}
\caption{Plots of $K_{3}$ as a function of $\tau$
with $N=40$, $\phi_{0}=0$, $\alpha=1/2$ and, $\omega=1$ for (a) $\Gamma=0.005$, (b) $\Gamma=0.1$.
The value $K_{3}$ converges to zero as $\tau$ becomes larger in (b).}
\label{figure-05ab}
\end{minipage}
\hspace{0.04\columnwidth}
\begin{minipage}[b]{0.45\linewidth}
\begin{center}
\includegraphics[keepaspectratio, scale=0.82]{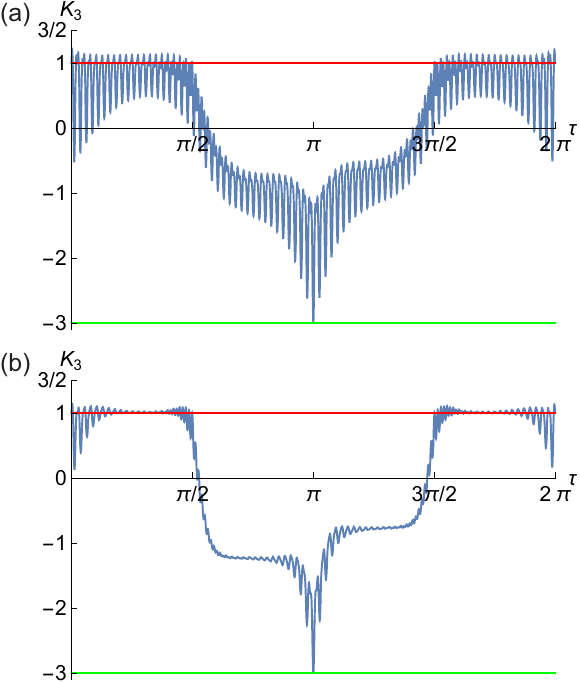}
\end{center}
\caption{Plots of $K_{3}$ as a function of $\tau$
with $N=40$, $\phi_{0}=0$, $\Gamma=0$ and, $\omega=1$ for (a) $\alpha=1$, (b) $\alpha=2$.}
\label{figure-06ab}
\end{minipage}
\end{figure}

Figures~\ref{figure-04ab}~(a), (b)
show plots of $K_{3}$ as functions of $\tau$
for $N=20$ and $40$, respectively.
Looking at these we note that oscillations of the graph increase as $N$ becomes larger.
Amplitudes of oscillations of short periods in the graph do not decrease even if $N$ becomes larger.
Thus, we can expect that a graph of $K_{3}$ as a function of $\tau$ contains oscillations of infinitely high frequency as $N\to\infty$.
Further Figs.~\ref{figure-05ab}~(a), (b)
show plots of $K_{3}$ as functions of $\tau$
for two different dissipation values, namely $\Gamma=0.005$ and $0.1$, respectively.
Looking at these plots we note that the amplitude of the oscillations of $K_{3}$ decreases as $\Gamma$ becomes larger.
More specifically looking at Fig.~\ref{figure-05ab}~(b),
we note that $K_{3}$ converges to zero as $\tau$ increases on condition $\Gamma>0$.

\begin{figure}[ht]
\centering
\begin{minipage}[b]{0.45\linewidth}
\begin{center}
\includegraphics[keepaspectratio, scale=0.82]{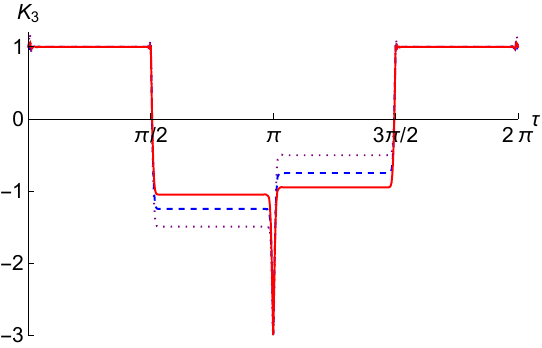}
\end{center}
\caption{Plots of $K_{3}$ as a function of $\tau$
for various $N$ with $\phi_{0}=0$, $\alpha=10$, $\Gamma=0$, and $\omega=1$.
The dotted purple, dashed blue, and solid red curves represent $N=100$, $200$, and $1000$, respectively.}
\label{figure-07}
\end{minipage}
\hspace{0.04\columnwidth}
\begin{minipage}[b]{0.45\linewidth}
\begin{center}
\includegraphics[keepaspectratio, scale=0.82]{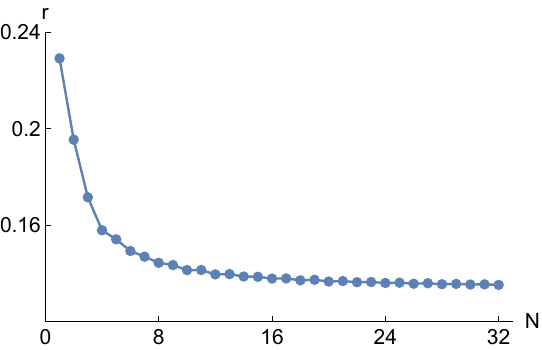}
\end{center}
\caption{Plot of the proportion $r$
as a function of $N$ for $\phi_{0}=0$, $\alpha=1/2$, $\Gamma=0$, and $\omega=1$.}
\label{figure-08}
\end{minipage}
\end{figure}

Next Figs.~\ref{figure-06ab}~(a), (b)
shows plots of $K_{3}$ as functions of $\tau$ for
$\alpha=1$ and $2$, respectively.
Figure~\ref{figure-07} shows plots of $K_{3}$ as a function of $\tau$
for $N=100$, $200$, and $1000$.
As $N$ becomes larger, the curves approach a step function
\begin{equation}
f(\tau)
=
\left\{
\begin{array}{ll}
1 & \mbox{for $0\leq\tau<\pi/2$ and $3\pi/2\leq\tau<2\pi$} \\
-1 & \mbox{for $\pi/2\leq\tau<3\pi/2$} \\
\end{array}
\right..
\label{step-function-01}
\end{equation}

We can now compute the proportion of the time cycle where the LGI is violated.
This we term the $r$ value and it is calculated as follows.
Starting with the dissipationless case, the period of $K_{3}$ as a function of $\tau$ is equal to $2\pi/\omega$.
We can then divide this period into $M$ slices of size $\Delta\tau=2\pi/(M\omega)$.
Within each slice we examine whether or not the LGI is violated at each $\tau_{m}=m\Delta\tau$ for $m=0,1,2,...,M$.
We count the number of slices $\tilde{M}$ where the LGI is violated.
Strictly speaking $0\leq\tilde{M}\leq M$.
Finally, we define the dimensionless ratio $r$ as
\begin{equation}
r=\tilde{M}/M.
\end{equation}
This ratio was defined for the dissipationless case but can also be used when dissipation is present.
In this case we set
$\Delta\tau=2\pi/(M\omega)$,
and can determine $r$ as previously described.
We can regard $r$ as the proportion of the LGI violation and explore its behavior.

\begin{figure}[ht]
\centering
\begin{minipage}[b]{0.45\linewidth}
\begin{center}
\includegraphics[keepaspectratio, scale=0.82]{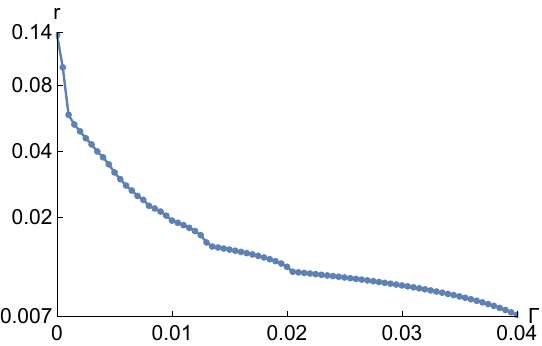}
\end{center}
\caption{Plot of the proportion $r$
as a function of $\Gamma$ for $\phi_{0}=0$, $\alpha=1/2$, $N=25$, and $\omega=1$.
The proportion $r$ is displayed on a logarithmic scale.}
\label{figure-09}
\end{minipage}
\hspace{0.04\columnwidth}
\begin{minipage}[b]{0.45\linewidth}
\begin{center}
\includegraphics[keepaspectratio, scale=0.79]{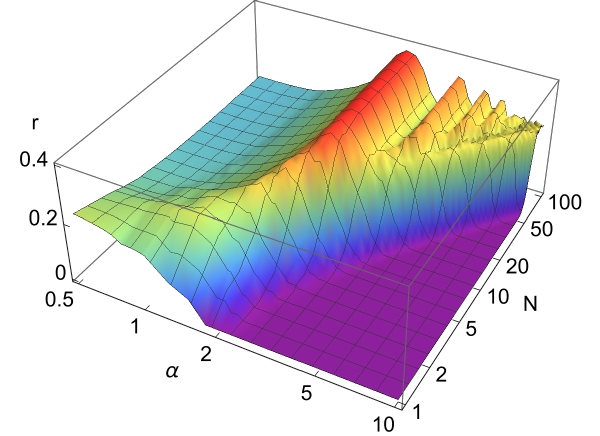}
\end{center}
\caption{Plot of the proportion $r$ as a function of real $\alpha$ and $N$
for $\phi_{0}=0$, $\Gamma=0$, and $\omega=1$.
The axes of $\alpha$ and $N$ are displayed on a logarithmic scale.}
\label{figure-10}
\end{minipage}
\end{figure}

Figure~\ref{figure-08} shows a plot of $r$ as a function of $N$.
The period of $K_{3}$ for $\tau$ is equal to $2\pi$
and we divide the period into $M=10^{6}$ slices of size $2\pi/M$.
Looking at Fig.~\ref{figure-08}, we note that $r$ approaches $0.137$ as $N$ becomes larger.
The graph of Fig.~\ref{figure-08} suggests that the violation is preserved as $N$ increases
in the dissipationless case.
However, if $\Gamma$ is not equal to zero we observe as seen in Fig.~\ref{figure-09},
the violation is getting smaller and $r$ approaches zero exponentially.

Next Fig.~\ref{figure-10} shows a three-dimensional plot of $r$
as a function of both $\alpha$ and $N$.
Here we assume $\alpha$ is real.
We observe from the plot that $r$ does not approach zero in a large $\alpha$ and $N$ limit.
In the dissipationless case the violation of the LGI does not disappear as $N$ increases even if $\alpha$ is large enough.
If we let $|\alpha|$ and $N$ become large the graph of $K_{3}$ can be approximated by the step function of Eq.~(\ref{step-function-01})
implying the violation of the LGI remains.

In Fig.~\ref{figure-10}, we have shown that the LGI of the coherent state $|\alpha\rangle$ reveals the violation in a large $N$ limit
even if we let $|\alpha|$ be large.
Because
\begin{equation}
\frac{2\Delta n\Delta\phi}{|\langle\alpha|[\hat{n},\hat{\phi}]|\alpha\rangle|}
\simeq
1
\end{equation}
holds for $|\alpha|\gg 1$
where $\hat{n}$ is the number operator of bosons
and
$\hat{\phi}$ is the Pegg--Barnett phase operator
defined in Eq.~(\ref{Pegg-Barnett-phase-operator-0})
with $s\to\infty$,
the coherent state $|\alpha\rangle$ shows the minimum uncertainty
\cite{Barnett1997,Buzek1992}.
However, Fig.~\ref{figure-10} tells us that large $|\alpha|$ reveals the violation of the LGI.
Thus,
$|\alpha\rangle$ keeps its nonclassical properties and it cannot be described with classical theory.

Next as shown in Figs.~\ref{figure-05ab}~(b) and \ref{figure-09},
if we introduce dissipation,
$K_{3}$ approaches zero as $\tau$ becomes larger and $r$ decreases exponentially
as $\Gamma$ becomes larger.
It was reported that quantum properties of a damped two-level system decay exponentially in time
because of the coupling to the thermal reservoir \cite{Friedenberger2017}.
In the current paper, we assume that there is no thermal fluctuation of the external reservoir
but the boson system suffers from dissipation due to the interaction with the zero-temperature environment.
In this scenario, we show the exponential decay of the LGI.
Thus, we can conclude that the interaction with the external system causes the reduction of quantumness,
for example, the violation of the LGI,
for the system.

The shape of the graph in Fig.~\ref{figure-04ab} (a) is very different from that in Fig.~\ref{figure-03ab} (a).
We can recognize the plot of Fig.~\ref{figure-04ab} (a) as an isolated one because Fig.~\ref{figure-03ab} (a) is common
for the graphs of the LGI, for example, as shown in Ref.~\cite{Emary2014}.
It surprises us that difference of observables $\hat{O}^{(\mbox{\scriptsize parity})}_{s}$ and $\hat{O}^{(+,-)}_{s}$
generates the difference between Figs.~\ref{figure-03ab} (a) and \ref{figure-04ab} (a).
Moreover, the graph of Fig.~\ref{figure-04ab} (a) changes into the step function shown in Figs.~\ref{figure-06ab} and ~\ref{figure-07}
under the large $|\alpha|$ and $N$ limit.
Thus, the choice of $\hat{O}^{(+,-)}_{s}$ for the LGI provides us with plenty of physical meanings.

In Fig.~\ref{figure-10} we evaluated the proportion $r$ in the range $0.5\leq\alpha\leq 10$ with $1\leq N\leq 100$.
By letting $\alpha$ and $N$ become larger,
we witness an oscillation of $r$.
This phenomenon remains to be investigated more precisely in the future.

\section{\label{section-discussions}conclusions and discussions}
As mentioned in our introduction, a violation of the LGI shows that there is a difference between quantum mechanics and macroscopic local realism.
Microscopic systems obviously tend to have a quantum nature and as such it is thus important to explore systems involving large particle/photon numbers.
Boson systems are interesting in that respect as one can easily create quantum states of light with varying mean photon number.
The simplest of course is the coherent state $|\alpha\rangle$ which is often considered the most ``classical'' state of light in nature for large $\alpha$.
As a continuous variable system, one needs to carefully consider the observables that will be used to determine whether the LGI is violated or not.
Here we used the discrete Pegg--Barnett phase operator with two different binning strategies to define an observable with $\pm 1$ values.
These observables were then used to determine the two-time correlation functions necessary to test the LGI.

In the ideal situation without dissipation, we showed that the coherent state $|\alpha\rangle$ violates the LGIs over a wide parameter regime
even with significant time difference $\tau$ between the measurements used to determine the two-time correlation functions.
In fact, we were able to show a LGI violation in the large $N$ limit even if we let $|\alpha|$ be large.
The coherent state $|\alpha\rangle$ keeps its nonclassical properties and it cannot be described with classical theory.
Further we also considered the effect of imperfect gate operations within our measurement devices and found the violation is maintained
in a reduced fashion over a wide range of parameters.
However, including dissipation within our model, we found as expected that the LGI violation does disappear as $\tau$ becomes large.
Regimes do however still exist where a violation can be seen.

\section*{Acknowledgments}
This work was supported by MEXT Quantum Leap Flagship Program (MEXT Q-LEAP) Grant Number JPMXS0120351339.

\appendix

\section{\label{Appendix-A}Solutions of Eq.~(\ref{master-equation-0})}
Let us begin by assuming that the initial state can be expressed by a density matrix made up of the component
$v(0)=|\alpha\rangle\langle \beta|$,
where $|\alpha\rangle$ and $|\beta\rangle$ represent the coherent states
with complex amplitudes $\alpha$ and $\beta$.
According to the master equation (\ref{master-equation-0}), $v(0)$ evolves as
\begin{equation}
v(t)
=
\exp
\Bigl(
-\frac{1}{2}(|\alpha|^{2}+|\beta|^{2}-2\alpha\beta^{*})
\Bigl[1-\exp[-2\Gamma t]\Bigr]
\Bigr)
|\alpha\exp(-i\Omega t)\rangle\langle\beta\exp(-i\Omega t)|,
\label{time-evolution-alpha-beta-0}
\end{equation}
with $\Omega=\omega-i\Gamma$ \cite{Walls1994,Barnett1997,Azuma2021}.
It is useful to point out that the time evolution of the initial state $|\alpha\rangle$ is given
by $|\alpha\exp(-i\Omega t)\rangle$.

Next consider that a component of the density matrix of the initial state is given by
$v(0)=|n\rangle\langle m|$,
where $|n\rangle$ and $|m\rangle$ represent the $n$th and $m$th Fock states.
It is straightforward to show that $v(0)$ evolves as
\begin{equation}
v(t)
=
\exp[-\Gamma t(n+m)-i\omega(n-m)t]
\sum_{l=0}^{\mbox{\scriptsize min}(n,m)}
\Bigl[
\exp[2\Gamma t]-1
\Bigr]^{l}
\sqrt{
\left(
\begin{array}{c}
n \\
l \\
\end{array}
\right)
\left(
\begin{array}{c}
m \\
l \\
\end{array}
\right)
}
|n-l\rangle\langle m-l|.
\label{time-evolution-n-m-0}
\end{equation}

\section{\label{Appendix-B}Explicit forms of
$|\psi'_{\pm}\rangle_{s}$
for Eqs.~(\ref{psi-dash-pm-0}) and (\ref{imperfect-gates-P-sigma-0})}
According to Eqs.~(\ref{definition-noisy-QFT-0}) and (\ref{definition-alpha-s-coherent}),
the noisy QFT transforms $|\alpha\rangle_{s}$ into
\begin{equation}
|\psi\rangle_{s}
=
C_{s}(\alpha)
\frac{1}{2^{L/2}}
\sum_{x=0}^{2^{L}-1}
\sum_{y=0}^{2^{L}-1}
\frac{\alpha^{x}}{\sqrt{x!}}
\exp[i\frac{2\pi}{2^{L}}xy
+i
\sum_{p\in\{2,...,L\}}
\sum_{q\in\{1,...L-1\},p>q,}
\Delta_{pq}x_{p}y_{L-q+1}
]|y\rangle.
\end{equation}
Because we want to observe $\hat{O}^{(+,-)}_{s}$,
we should measure $|y_{L}\rangle$.
If we detect $y_{L}=0$, that is to say, $O^{(+,-)}_{s}=1$,
the wave function collapses from $|\psi\rangle_{s}$ to a non-normalized one,
\begin{equation}
|\psi_{+}\rangle_{s}
=
C_{s}(\alpha)
\frac{1}{2^{L/2}}
\sum_{x=0}^{2^{L}-1}
\sum_{y'=0}^{2^{L-1}-1}
\frac{\alpha^{x}}{\sqrt{x!}}
\exp[i\frac{2\pi}{2^{L}}xy'
+i
\sum_{p\in\{3,...,L\}}
\sum_{q\in\{2,...L-1\},p>q,}
\Delta_{pq}x_{p}y'_{L-q+1}
]|y'\rangle.
\label{psi-plus-0}
\end{equation}
In contrast,
if we detect $y_{L}=1$, that is to say, $O^{(+,-)}_{s}=-1$,
the wave function collapses from $|\psi\rangle_{s}$ to a non-normalized one,
\begin{eqnarray}
|\psi_{-}\rangle_{s}
&=&
C_{s}(\alpha)
\frac{1}{2^{L/2}}
\sum_{x=0}^{2^{L}-1}
\sum_{y'=0}^{2^{L-1}-1}
\frac{\alpha^{x}}{\sqrt{x!}}
\exp[i\frac{2\pi}{2^{L}}xy'
+i2\pi\frac{x}{2} \nonumber \\
&&
+i
\sum_{p\in\{3,...,L\}}
\sum_{q\in\{2,...L-1\},p>q,}
\Delta_{pq}x_{p}y'_{L-q+1}
+i
\sum_{p\in\{2,...,L\}}
\Delta_{p1}x_{p}
]|y'+2^{L-1}\rangle.
\label{psi-minus-0}
\end{eqnarray}
We can derive $||\psi_{+}\rangle_{s}|^{2}=||\psi_{-}\rangle_{s}|^{2}=1/2$ with ease.

To accomplish the QND measurement,
we need to apply noisy QFT${}^{-1}$ to $|\psi_{\pm}\rangle_{s}$
and obtain $|\psi'_{\pm}\rangle_{s}$.
From Eqs.~(\ref{psi-plus-0}) and (\ref{psi-minus-0}),
we write these explicitly as
\begin{eqnarray}
|\psi'_{+}\rangle_{s}
&=&
(\mbox{noisy QFT}^{-1})|\psi_{+}\rangle_{s} \nonumber \\
&=&
\frac{C_{s}(\alpha)}{d}
\sum_{x=0}^{d-1}
\sum_{y'=0}^{(d/2)-1}
\sum_{z=0}^{d-1}
\frac{\alpha^{x}}{\sqrt{x!}}
\exp
\Biggl[
i\frac{2\pi}{d}y'(x-z)
+i
\sum_{\substack{{p\in\{3,...,L\},}\\{q\in\{2,...L-1\},}\\{p>q}}}
\Delta_{pq}x_{p}y'_{L-q+1} \nonumber \\
&&
\quad\quad\quad\quad
\quad\quad\quad\quad
\quad\quad\quad\quad
\quad\quad\quad\quad
\quad\quad\quad\quad
\quad\quad\quad\quad
\quad
+i
\sum_{\substack{{p\in\{2,...,L-1\},}\\{q\in\{1,...L-2\},}\\{p>q}}}
\tilde{\Delta}_{pq}y'_{p}z_{L-q+1}
\Biggr]
|z\rangle, \nonumber \\
|\psi'_{-}\rangle_{s}
&=&
(\mbox{noisy QFT}^{-1})|\psi_{-}\rangle_{s} \nonumber \\
&=&
\frac{C_{s}(\alpha)}{d}
\sum_{x=0}^{d-1}
\sum_{y'=0}^{(d/2)-1}
\sum_{z=0}^{d-1}
\frac{\alpha^{x}}{\sqrt{x!}}
\exp
\Biggl[
i\frac{2\pi}{d}y'(x-z)
+i\pi (x-z)
+i
\sum_{\substack{{p\in\{3,...,L\},}\\{q\in\{2,...L-1\},}\\{p>q}}}
\Delta_{pq}x_{p}y'_{L-q+1} \nonumber \\
&&
\quad\quad\quad\quad
\quad\quad\quad\quad
\quad\quad\quad\quad
\quad\quad\quad\quad
+i
\sum_{p\in\{2,...,L\}}
\Delta_{p1}x_{p}
+i
\sum_{\substack{{p\in\{2,...,L-1\},}\\{q\in\{1,...L-2\},}\\{p>q}}}
\tilde{\Delta}_{pq}y'_{p}z_{L-q+1}
\Biggr]
|z\rangle, \nonumber \\
\end{eqnarray}
where $\{\Delta_{pq}\}$ and $\{\tilde{\Delta}_{pq}\}$ are Gaussian variables generated by the noisy QFT and $(\mbox{noisy QFT})^{-1}$, respectively.

\section{\label{Appendix-C}Derivation of Eqs.~(\ref{C21-parity-formula-0}) and (\ref{Trevenodd-L1L2-tau})}
We prepare the coherent state $|\alpha\rangle$ as our system's initial state at $t_{1}=0$.
The probability that we obtain $O_{1}=+1$ at $t_{1}=0$ is given by
\begin{eqnarray}
P_{1}(+1)
&=&
\langle\alpha|\Pi^{(\mbox{\scriptsize even})}_{s}|\alpha\rangle \nonumber \\
&=&
\sum_{m=0}^{N}|{}_{s}\langle\phi_{2m}|\alpha\rangle|^{2} \nonumber \\
&=&
(s+1)^{-1}\exp(-|\alpha|^{2})
\sum_{m=0}^{N}
\sum_{n=0}^{s}\sum_{n'=0}^{s}
\exp[-i(n-n')\phi_{2m}]
\tilde{\alpha}(n)
\tilde{\alpha}^{*}(n'),
\label{P1_plus1-00}
\end{eqnarray}
where
$
\tilde{\alpha}(n)
=
\alpha^{n}/\sqrt{n!}
$,
and
$
\tilde{\alpha}^{*}(n)
=
(\alpha^{*})^{n}/\sqrt{n!}
$.
Here, we examine the following finite series:
\begin{eqnarray}
S_{1}(N,n-n')
&=&
\sum_{m=0}^{N}
\exp[-i(n-n')\phi_{2m}] \nonumber \\
&=&
\exp[-i(n-n')\phi_{0}]
\frac{\exp[-2i(n-n')\pi]-1}{\exp[-2i(n-n')\pi/(N+1)]-1}.
\end{eqnarray}
To evaluate the above series, we draw attention to the fact $-2N-1\leq n-n'\leq 2N+1$.
Then, we obtain
\begin{equation}
S_{1}(N,n-n')
=
\left\{
\begin{array}{ll}
N+1 & \mbox{for $n=n'$} \\
(N+1)\exp[-i(N+1)\phi_{0}] & \mbox{for $n-n'=N+1$} \\
(N+1)\exp[i(N+1)\phi_{0}] & \mbox{for $n-n'=-N-1$} \\
0 & \mbox{for others}
\end{array}
\right..
\label{formula-S1-01}
\end{equation}
Based on the above calculations, we obtain $P_{1}(+1)$ as
\begin{eqnarray}
P_{1}(+1)
&=&
\frac{1}{2}\exp(-|\alpha|^{2})
\{
\sum_{n=0}^{s}|\tilde{\alpha}(n)|^{2}
+
\exp[-i(N+1)\phi_{0}]
\sum_{n=N+1}^{s}
\tilde{\alpha}(n)
\tilde{\alpha}^{*}(n-N-1) \nonumber \\
&&
+
\exp[i(N+1)\phi_{0}]
\sum_{n=0}^{N}
\tilde{\alpha}(n)
\tilde{\alpha}^{*}(n+N+1)
\}.
\label{P1_plus1-01}
\end{eqnarray}
Changing the representation of $P_{1}(+1)$ from Eq.~(\ref{P1_plus1-00}) to Eq.~(\ref{P1_plus1-01}),
we remove the factor $(s+1)^{-1}$ because $s=2N+1$
and
$S_{1}(N,n-n')$ includes the factor $(N+1)$ as shown in Eq.~(\ref{formula-S1-01}).
Thus, the operation of taking large $s$ limit does not causes any troubles.

If we observe $\Pi^{(\mbox{\scriptsize even})}_{s}$ at $t_{1}=0$, the wave function collapses projecting us to the non-normalized sate,
\begin{eqnarray}
\Pi^{(\mbox{\scriptsize even})}_{s}|\alpha\rangle
&=&
\sum_{m=0}^{N}{}_{s}\langle\phi_{2m}|\alpha\rangle|\phi_{2m}\rangle_{s} \nonumber \\
&=&
(s+1)^{-1}\exp(-|\alpha|^{2}/2)
\sum_{n=0}^{s}\sum_{n'=0}^{s}
\tilde{\alpha}(n)
\sum_{m=0}^{N}
\exp[-i(n-n')\phi_{2m}]|n'\rangle \nonumber \\
&=&
\frac{1}{2}\exp(-|\alpha|^{2}/2)
\{
\sum_{n=0}^{s}\tilde{\alpha}(n)|n\rangle
+
\exp[-i(N+1)\phi_{0}]
\sum_{n=0}^{N}
\tilde{\alpha}(n+N+1)|n\rangle \nonumber \\
&&
+
\exp[i(N+1)\phi_{0}]
\sum_{n=N+1}^{s}
\tilde{\alpha}(n-N-1)|n\rangle
\}.
\label{Pi_even_s_alpha-01}
\end{eqnarray}
In the right-hand side of the above equation,
the factor $(s+1)^{-1}$ disappears because of the same reason mentioned in Eq.~(\ref{P1_plus1-01}).

Smilarly, we obtain the probability $P_{1}(-1)$ if we observe $O_{1}=-1$ at $t_{1}=0$
and the sate after the wave-function collapse $\Pi^{(\mbox{\scriptsize odd})}_{s}|\alpha\rangle$.
In these derivations, the following formula is useful:
\begin{eqnarray}
S_{2}(N,n-n')
&=&
\sum_{m=0}^{N}
\exp[-i(n-n')\phi_{2m+1}] \nonumber \\
&=&
\exp[-i(n-n')\phi_{0}-i(n-n')\frac{2\pi}{s+1}]
\frac{\exp[-2i(n-n')\pi]-1}{\exp[-2i(n-n')\pi/(N+1)]-1} \nonumber \\
&=&
\left\{
\begin{array}{ll}
N+1 & \mbox{for $n=n'$} \\
-(N+1)\exp[-i(N+1)\phi_{0}] & \mbox{for $n-n'=N+1$} \\
-(N+1)\exp[i(N+1)\phi_{0}] & \mbox{for $n-n'=-N-1$} \\
0 & \mbox{for others} \\
\end{array}
\right..
\end{eqnarray}

From the above formula, we obtain
\begin{eqnarray}
P_{1}(-1)
&=&
\frac{1}{2}\exp(-|\alpha|^{2})
\{
\sum_{n=0}^{s}|\tilde{\alpha}(n)|^{2}
-
\exp[-i(N+1)\phi_{0}]
\sum_{n=N+1}^{s}
\tilde{\alpha}(n)\tilde{\alpha}^{*}(n-N-1) \nonumber \\
&&
-
\exp[i(N+1)\phi_{0}]
\sum_{n=0}^{N}
\tilde{\alpha}(n)\tilde{\alpha}^{*}(n+N+1)
\},
\end{eqnarray}
\begin{eqnarray}
\Pi^{(\mbox{\scriptsize odd})}_{s}|\alpha\rangle
&=&
\frac{1}{2}\exp(-|\alpha|^{2}/2)
\{
\sum_{n=0}^{s}\tilde{\alpha}(n)|n\rangle
-
\exp[-i(N+1)\phi_{0}]
\sum_{n=0}^{N}
\tilde{\alpha}(n+N+1)|n\rangle \nonumber \\
&&
-
\exp[i(N+1)\phi_{0}]
\sum_{n=N+1}^{s}
\tilde{\alpha}(n-N-1)|n\rangle
\}.
\end{eqnarray}

It is useful to describe a non-normalized state obtained by the wave-function collapse
for the observation of $O_{1}=\pm 1$ at $t_{1}=0$ as $w_{1\pm}(0)$.
This enables us to write $w_{1\pm}(0)$ explicitly as follows:
\begin{eqnarray}
w_{1\pm}(0)
&=&
\Pi^{(\chi)}_{s}|\alpha\rangle\langle\alpha|\Pi^{(\chi)}_{s} \nonumber \\
&=&
\frac{1}{4}
\exp[-|\alpha|^{2}]
[
K_{1}(0)+K_{2}(0)+K_{3}(0)
\pm L_{1}(0)
\pm L_{2}(0)
+L_{3}(0)
\pm L_{1}^{\dagger}(0)
\pm L_{2}^{\dagger}(0)
+L_{3}^{\dagger}(0)
],
\end{eqnarray}
where $\chi=\mbox{even}$ for $w_{1+}(0)$
and $\mbox{odd}$ for $w_{1-}(0)$.
Further
\begin{eqnarray}
K_{1}(0)
&=&
\sum_{n=0}^{s}\sum_{n'=0}^{s}
\tilde{\alpha}(n)
\tilde{\alpha}^{*}(n')
|n\rangle\langle n'|, \nonumber \\
K_{2}(0)
&=&
\sum_{n=0}^{N}\sum_{n'=0}^{N}
\tilde{\alpha}(n+N+1)
\tilde{\alpha}^{*}(n'+N+1)
|n\rangle\langle n'|, \nonumber \\
K_{3}(0)
&=&
\sum_{n=N+1}^{s}\sum_{n'=N+1}^{s}
\tilde{\alpha}(n-N-1)
\tilde{\alpha}^{*}(n'-N-1)
|n\rangle\langle n'|, \nonumber \\
L_{1}(0)
&=&
\exp[i(N+1)\phi_{0}]
\sum_{n=0}^{s}
\sum_{n'=0}^{N}
\tilde{\alpha}(n)
\tilde{\alpha}^{*}(n'+N+1)
|n\rangle\langle n'|, \nonumber \\
L_{2}(0)
&=&
\exp[-i(N+1)\phi_{0}]
\sum_{n=0}^{s}
\sum_{n'=N+1}^{s}
\tilde{\alpha}(n)
\tilde{\alpha}^{*}(n'-N-1)
|n\rangle\langle n'|, \nonumber \\
L_{3}(0)
&=&
\exp[-2i(N+1)\phi_{0}]
\sum_{n=0}^{N}
\sum_{n'=N+1}^{s}
\tilde{\alpha}(n+N+1)
\tilde{\alpha}^{*}(n'-N-1)
|n\rangle\langle n'|.
\label{KL-parity}
\end{eqnarray}

We can now consider the time evolution of the state $w_{1\pm}(0)$ from $t_{1}=0$ until $t_{2}=\tau$.
At time $\tau$ the state has the form
\begin{equation}
w_{1\pm}(\tau)
=
\frac{1}{4}
\exp[-|\alpha|^{2}]
[
K_{1}(\tau)+K_{2}(\tau)+K_{3}(\tau)
\pm L_{1}(\tau)
\pm L_{2}(\tau)
+L_{3}(\tau)
\pm L_{1}^{\dagger}(\tau)
\pm L_{2}^{\dagger}(\tau)
+L_{3}^{\dagger}(\tau)
].
\end{equation}
We consider the explicit forms of $K_{1}(\tau)$, $K_{2}(\tau)$, $K_{3}(\tau)$, $L_{1}(\tau)$, $L_{2}(\tau)$, and $L_{3}(\tau)$ later.

Now we write the probability that we observe $O_{1}=+1$ and $O_{2}=\pm 1$ at $t_{1}=0$ and $t_{2}=\tau$
as $p_{1+,2\pm}$, respectively where
\begin{eqnarray}
p_{1+,2\pm}
&=&
\mbox{Tr}[\Pi^{(\chi)}_{s}w_{1+}(\tau)] \nonumber \\
&=&
\frac{1}{4}\exp[-|\alpha|^{2}]
\mbox{Tr}
[\Pi^{(\chi)}_{s}
\{
K_{1}(\tau)+K_{2}(\tau)+K_{3}(\tau) \nonumber \\
&&
+
L_{1}(\tau)+L_{2}(\tau)+L_{3}(\tau)
+
L_{1}^{\dagger}(\tau)+L_{2}^{\dagger}(\tau)+L_{3}^{\dagger}(\tau)
\}
],
\end{eqnarray}
with $\chi=\mbox{even}$ for $p_{1+,2+}$ and $\mbox{odd}$ for $p_{1+,2-}$.
Similarly, we write the probability that we observe $O_{1}=-1$ and $O_{2}=\pm 1$ at $t_{1}=0$ and $t_{2}=\tau$
as $p_{1-,2\pm}$ respectively as
\begin{eqnarray}
p_{1-,2\pm}
&=&
\mbox{Tr}[\Pi^{(\chi)}_{s}w_{1-}(\tau)] \nonumber \\
&=&
\frac{1}{4}\exp[-|\alpha|^{2}]
\mbox{Tr}
[\Pi^{(\chi)}_{s}
\{
K_{1}(\tau)+K_{2}(\tau)+K_{3}(\tau) \nonumber \\
&&
-
L_{1}(\tau)-L_{2}(\tau)+L_{3}(\tau)
-
L_{1}^{\dagger}(\tau)-L_{2}^{\dagger}(\tau)+L_{3}^{\dagger}(\tau)
\}
],
\end{eqnarray}
where $\chi=\mbox{even}$ for $p_{1-,2+}$ and $\mbox{odd}$ for $p_{1-,2-}$.

Next we can write the correlation function $C_{12}$ as
\begin{equation}
C_{12}
=
p_{1+,2+}
-
p_{1+,2-}
-
p_{1-,2+}
+
p_{1-,2-},
\label{C21-parity-formula-1}
\end{equation}
allowing us to obtain Eq.~(\ref{C21-parity-formula-0}).
In the computation of $C_{12}$ according to Eq.~(\ref{C21-parity-formula-1}),
$\mbox{Tr}[\Pi_{s}^{(\chi)}(L_{1}(\tau)+L_{2}(\tau)+L_{1}^{\dagger}(\tau)+L_{2}^{\dagger}(\tau))]
=
2\mbox{Re}\{\mbox{Tr}[\Pi_{s}^{(\chi)}(L_{1}(\tau)+L_{2}(\tau))]\}$
and this is the reason why Eq.~(\ref{C21-parity-formula-0}) includes only the real part of Eq.~(\ref{Trevenodd-L1L2-tau}).
Because of Eq.~(\ref{C21-parity-formula-0}), we only need to focus on the computation of
$\mbox{Tr}[\Pi^{(\chi)}_{s}(L_{1}(\tau)+L_{2}(\tau))]$.

From Eqs.~(\ref{time-evolution-n-m-0}) and (\ref{KL-parity}),
we can express $L_{1}(\tau)$ and $L_{2}(\tau)$ as
\begin{eqnarray}
L_{1}(\tau)
&=&
\exp[i(N+1)\phi_{0}]
\sum_{n=0}^{s}
\sum_{n'=0}^{N}
\tilde{\alpha}(n)
\tilde{\alpha}^{*}(n'+N+1)
\exp[-\Gamma \tau(n+n')-i\omega(n-n')\tau] \nonumber \\
&&
\quad\quad\quad\quad
\quad\quad\quad\quad
\quad\quad\quad\quad
\quad\quad\quad\quad
\times
\sum_{l=0}^{\mbox{\scriptsize min}(n,n')}
\Bigl[
\exp[2\Gamma t]-1
\Bigr]^{l}
\sqrt{
\left(
\begin{array}{c}
n \\
l \\
\end{array}
\right)
\left(
\begin{array}{c}
n' \\
l \\
\end{array}
\right)
}
|n-l\rangle\langle n'-l|,
\label{L1-tau-parity}
\end{eqnarray}
\begin{eqnarray}
L_{2}(\tau)
&=&
\exp[-i(N+1)\phi_{0}]
\sum_{n=0}^{s}
\sum_{n'=N+1}^{s}
\tilde{\alpha}(n)
\tilde{\alpha}^{*}(n'-N-1)
\exp[-\Gamma \tau(n+n')-i\omega(n-n')\tau] \nonumber \\
&&
\quad\quad\quad\quad
\quad\quad\quad\quad
\quad\quad\quad\quad
\quad\quad\quad\quad
\times
\sum_{l=0}^{\mbox{\scriptsize min}(n,n')}
\Bigl[
\exp[2\Gamma t]-1
\Bigr]^{l}
\sqrt{
\left(
\begin{array}{c}
n \\
l \\
\end{array}
\right)
\left(
\begin{array}{c}
n' \\
l \\
\end{array}
\right)
}
|n-l\rangle\langle n'-l|.
\label{L2-tau-parity}
\end{eqnarray}

Thus, we can describe $\mbox{Tr}[\Pi^{(\chi)}_{s}L_{1}(\tau)]$ and $\mbox{Tr}[\Pi^{(\chi)}_{s}L_{2}(\tau)]$
for $\chi\in\{\mbox{even},\mbox{odd}\}$
explicitly in the form,
\begin{eqnarray}
\mbox{Tr}[\Pi^{(\chi)}_{s}L_{1}(\tau)]
&=&
\frac{1}{2}\exp[i(N+1)\phi_{0}]
\sum_{n=0}^{N}
\tilde{\alpha}(n)
\tilde{\alpha}^{*}(n+N+1) \nonumber \\
&&
\pm
\frac{1}{2}
\sum_{n=0}^{N}
|\tilde{\alpha}(n+N+1)|^{2}
\exp[-\Gamma \tau(2n+N+1)-i\omega(N+1)\tau] \nonumber \\
&&
\times
\sum_{l=0}^{n}
[\exp(2\Gamma t)-1]^{l}
\sqrt{
\left(
\begin{array}{c}
n+N+1 \\
l \\
\end{array}
\right)
\left(
\begin{array}{c}
n \\
l \\
\end{array}
\right)
},
\label{TrPevenodd-L1tau-0}
\end{eqnarray}
\begin{eqnarray}
\mbox{Tr}[\Pi^{(\chi)}_{s}L_{2}(\tau)]
&=&
\frac{1}{2}\exp[-i(N+1)\phi_{0}]
\sum_{n=0}^{N}
\tilde{\alpha}(n+N+1)
\tilde{\alpha}^{*}(n) \nonumber \\
&&
\pm
\frac{1}{2}
\sum_{n=0}^{N}
|\tilde{\alpha}(n)|^{2}
\exp[-\Gamma \tau(2n+N+1)+i\omega(N+1)\tau] \nonumber \\
&&
\times
\sum_{l=0}^{n}
[\exp(2\Gamma t)-1]^{l}
\sqrt{
\left(
\begin{array}{c}
n+N+1 \\
l \\
\end{array}
\right)
\left(
\begin{array}{c}
n \\
l \\
\end{array}
\right)
},
\label{TrPevenodd-L2tau-0}
\end{eqnarray}
where the $\pm$ symbol is replaced by the plus sign for $\chi=\mbox{even}$ and replaced by the minus sign for $\chi=\mbox{odd}$.
It is now straightforward to derive Eq.~(\ref{Trevenodd-L1L2-tau}).

\section{\label{Appendix-D}Derivation of Eqs.~(\ref{C21-formula-0}) and (\ref{Trpm-L-tau})}
We prepare the coherent state $|\alpha\rangle$ as our system's initial state at $t_{1}=0$.
The probability that we obtain $O_{1}=+1$ at $t_{1}=0$ is given by
\begin{eqnarray}
P_{1}(+1)
&=&
\langle\alpha|\Pi^{(+)}_{s}|\alpha\rangle \nonumber \\
&=&
\sum_{m=0}^{N}|{}_{s}\langle\phi_{m}|\alpha\rangle|^{2} \nonumber \\
&=&
(s+1)^{-1}\exp(-|\alpha|^{2})
\sum_{m=0}^{N}
\sum_{n=0}^{s}\sum_{n'=0}^{s}
\exp[-i(n-n')\phi_{m}]
\tilde{\alpha}(n)
\tilde{\alpha}^{*}(n').
\label{P1-plus1-02}
\end{eqnarray}
Here, we examine the following finite series:
\begin{eqnarray}
S_{3}(N,n-n')
&=&
\sum_{m=0}^{N}
\exp[-i(n-n')\phi_{m}] \nonumber \\
&=&
\exp[-i(n-n')\phi_{0}]
\frac{\exp[-i(n-n')\pi]-1}{\exp[-i(n-n')\pi/(N+1)]-1}.
\end{eqnarray}
To evaluate the above series, we draw attention to the fact $-2N-1\leq n-n'\leq 2N+1$.
Then, we obtain
\begin{equation}
S_{3}(N,n-n')
=
\left\{
\begin{array}{ll}
N+1 & \mbox{for $n=n'$} \\
0 & \mbox{for $(n-n')$ is even and $n\neq n'$} \\
\begin{array}{l}
-2\exp[-i(n-n')\phi_{0}] \\
/\{\exp[-i(n-n')\pi/(N+1)]-1\} \\
\end{array}
& \mbox{for $(n-n')$ is odd} \\
\end{array}
\right..
\label{formula-S3-01}
\end{equation}
Based on the above calculations, we obtain $P_{1}(+1)$ as
\begin{eqnarray}
P_{1}(+1)
&=&
\frac{1}{2}\exp(-|\alpha|^{2})\sum_{n=0}^{s}|\tilde{\alpha}(n)|^{2} \nonumber \\
&&
+
(s+1)^{-1}\exp(-|\alpha|^{2})
\sum_{n=0}^{s}
\sum_{n'=0, n-n'=\mbox{\scriptsize odd}}^{s}
(-2)\frac{\exp[-i(n-n')\phi_{0}]}{\exp[-i(n-n')\pi/(N+1)]-1}
\tilde{\alpha}(n)
\tilde{\alpha}^{*}(n').
\label{P1-plus1-03}
\end{eqnarray}
Changing the representation of $P_{1}(+1)$ from Eq.~(\ref{P1-plus1-02}) to Eq.~(\ref{P1-plus1-03}),
we eliminate the factor $(s+1)^{-1}$
because $S_{3}(N,n-n')$ includes the factor $(N+1)$ as shown in Eq.~(\ref{formula-S3-01}).
In the second term of the right-hand side of Eq.~(\ref{P1-plus1-03}),
we obtain
$\exp[-i(n-n')\pi/(N+1)]-1
\simeq
-i(n-n')\pi/(N+1)$
for $N\gg 1$
and we can cancel the factor $(s+1)^{-1}$ and $(N+1)$ because of $s=2N+1$.
Thus, the operation of taking large $s$ limit does not cause any troubles when we evaluate $P_{1}(+1)$.
We can apply this technique
to Eqs.~(\ref{Pi-plus-s-01}), (\ref{P-1-minus-1-01}), (\ref{Pi-minus-s-01}), (\ref{KL-0}),
(\ref{L-tau}), (\ref{TrPplusminusLtau-0}), and (\ref{Trpm-L-tau}), as well.

If we observe $\Pi^{(+)}_{s}$ at $t_{1}=0$, our wave function collapses to the non-normalized sate,
\begin{eqnarray}
\Pi^{(+)}_{s}|\alpha\rangle
&=&
\sum_{m=0}^{N}{}_{s}\langle\phi_{m}|\alpha\rangle|\phi_{m}\rangle_{s} \nonumber \\
&=&
(s+1)^{-1}\exp(-|\alpha|^{2}/2)
\sum_{n=0}^{s}\sum_{n'=0}^{s}
\tilde{\alpha}(n)
\sum_{m=0}^{N}
\exp[-i(n-n')\phi_{m}]|n'\rangle \nonumber \\
&=&
\frac{1}{2}\exp(-|\alpha|^{2}/2)
\sum_{n=0}^{s}\tilde{\alpha}(n)|n\rangle \nonumber \\
&&
+
(s+1)^{-1}\exp(-|\alpha|^{2}/2)
\sum_{n=0}^{s}
\sum_{n'=0, n-n'=\mbox{\scriptsize odd}}^{s}
(-2)
\frac{\exp[-i(n-n')\phi_{0}]}{\exp[-i(n-n')\pi/(N+1)]-1}
\tilde{\alpha}(n)|n'\rangle.
\label{Pi-plus-s-01}
\end{eqnarray}

Similarly, we obtain the probability $P_{1}(-1)$ if we observe $O_{1}=-1$ at $t_{1}=0$
and the sate after the wave-function collapse $\Pi^{(-)}_{s}|\alpha\rangle$.
In these derivations, the following formula is useful:
\begin{eqnarray}
S_{4}(N,n-n')
&=&
\sum_{m=N+1}^{s}
\exp[-i(n-n')\phi_{m}] \nonumber \\
&=&
\exp[-i(n-n')\phi_{0}-i(n-n')\pi]
\frac{\exp[-i(n-n')\pi]-1}{\exp[-i(n-n')\pi/(N+1)]-1} \nonumber \\
&=&
\left\{
\begin{array}{ll}
N+1 & \mbox{for $n=n'$} \\
0 & \mbox{for $(n-n')$ is even and $n\neq n'$} \\
\begin{array}{l}
2\exp[-i(n-n')\phi_{0}] \\
/\{\exp[-i(n-n')\pi/(N+1)]-1\} \\
\end{array}
& \mbox{for $(n-n')$ is odd} \\
\end{array}
\right..
\end{eqnarray}

From the above formula, we obtain
\begin{eqnarray}
P_{1}(-1)
&=&
\frac{1}{2}\exp(-|\alpha|^{2})\sum_{n=0}^{s}|\tilde{\alpha}(n)|^{2} \nonumber \\
&&
-
(s+1)^{-1}\exp(-|\alpha|^{2})
\sum_{n=0}^{s}
\sum_{n'=0, n-n'=\mbox{\scriptsize odd}}^{s}
(-2)\frac{\exp[-i(n-n')\phi_{0}]}{\exp[-i(n-n')\pi/(N+1)]-1}
\tilde{\alpha}(n)
\tilde{\alpha}^{*}(n'),
\label{P-1-minus-1-01}
\end{eqnarray}
\begin{eqnarray}
\Pi^{(-)}_{s}|\alpha\rangle
&=&
\frac{1}{2}\exp(-|\alpha|^{2}/2)
\sum_{n=0}^{s}\tilde{\alpha}(n)|n\rangle \nonumber \\
&&
-
(s+1)^{-1}\exp(-|\alpha|^{2}/2)
\sum_{n=0}^{s}
\sum_{n'=0, n-n'=\mbox{\scriptsize odd}}^{s}
(-2)
\frac{\exp[-i(n-n')\phi_{0}]}{\exp[-i(n-n')\pi/(N+1)]-1}
\tilde{\alpha}(n)|n'\rangle.
\label{Pi-minus-s-01}
\end{eqnarray}

We can now describe the non-normalized state obtained by the wave-function collapse
after the observation of $O_{1}=\pm 1$ at $t_{1}=0$ as
\begin{eqnarray}
\lefteqn{w_{1\pm}(0)} \nonumber \\
&=&
\Pi^{(\pm)}_{s}|\alpha\rangle\langle\alpha|\Pi^{(\pm)}_{s} \nonumber \\
&=&
\exp[-|\alpha|^{2}]
\{
K_{1}(0)
\pm
[L(0)+L^{\dagger}(0)]
+
K_{2}(0)
\}, \nonumber \\
\end{eqnarray}
where
\begin{eqnarray}
K_{1}(0)
&=&
\frac{1}{4}
\sum_{n=0}^{s}\sum_{n'=0}^{s}
\tilde{\alpha}(n)
\tilde{\alpha}^{*}(n')
|n\rangle\langle n'|, \nonumber \\
K_{2}(0)
&=&
(s+1)^{-2}
\sum_{n=0}^{s}
\sum_{\substack{{n'=0,}\\{n-n'=\mbox{\scriptsize odd}}}}^{s}
\sum_{m=0}^{s}
\sum_{\substack{{m'=0,}\\{m-m'=\mbox{\scriptsize odd}}}}^{s} \nonumber \\
&&
\times
4
\tilde{\alpha}(n)
\tilde{\alpha}^{*}(m)
\frac{\exp[-i(n-n')\phi_{0}]}{\exp[-i(n-n')\pi/(N+1)]-1}
\frac{\exp[i(m-m')\phi_{0}]}{\exp[i(m-m')\pi/(N+1)]-1}
|n'\rangle\langle m'|, \nonumber \\
L(0)
&=&
\frac{1}{2}(s+1)^{-1}
\sum_{n=0}^{s}
\sum_{m=0}^{s}
\sum_{\substack{{m'=0,}\\{m-m'=\mbox{\scriptsize odd}}}}^{s}
\tilde{\alpha}(n)
\tilde{\alpha}^{*}(m)
(-2)
\frac{\exp[i(m-m')\phi_{0}]}{\exp[i(m-m')\pi/(N+1)]-1}
|n\rangle\langle m'|. \nonumber \\
\label{KL-0}
\end{eqnarray}

This enables us now to consider the time evolution of the state $w_{1\pm}(0)$ from $t_{1}=0$ until $t_{2}=\tau$.
The resulting state has the form
\begin{equation}
w_{1\pm}(\tau)
=
\exp[-|\alpha|^{2}]
\{
K_{1}(\tau)
\pm
[L(\tau)+L^{\dagger}(\tau)]
+
K_{2}(\tau)
\}.
\end{equation}
Let us now describe the probability that we observe $O_{1}=+1$ and $O_{2}=\pm 1$ at $t_{1}=0$ and $t_{2}=\tau$
respectively as
\begin{eqnarray}
p_{1+,2\pm}
&=&
\mbox{Tr}[\Pi^{(\pm)}_{s}w_{1+}(\tau)] \nonumber \\
&=&
\exp[-|\alpha|^{2}]
\mbox{Tr}
\Bigl[
\Pi^{(\pm)}_{s}
[
K_{1}(\tau)
+
(L(\tau)+L^{\dagger}(\tau))
+
K_{2}(\tau)
]
\Bigr].
\end{eqnarray}
Similarly, we write the probability that we observe $O_{1}=-1$ and $O_{2}=\pm 1$ at $t_{1}=0$ and $t_{2}=\tau$ respectively
as
\begin{eqnarray}
p_{1-,2\pm}
&=&
\mbox{Tr}[\Pi^{(\pm)}_{s}w_{1-}(\tau)] \nonumber \\
&=&
\exp[-|\alpha|^{2}]
\mbox{Tr}
\Bigl[
\Pi^{(\pm)}_{s}
[
K_{1}(\tau)
-
(L(\tau)+L^{\dagger}(\tau))
+
K_{2}(\tau)
]
\Bigr].
\end{eqnarray}
We can now write the correlation function $C_{12}$ as
\begin{equation}
C_{12}
=
p_{1+,2+}
-
p_{1+,2-}
-
p_{1-,2+}
+
p_{1-,2-},
\label{C21-formula-1}
\end{equation}
and so obtain Eq.~(\ref{C21-formula-0}).

From Eqs.~(\ref{time-evolution-n-m-0}) and (\ref{KL-0}),
we can express $L(\tau)$ as
\begin{eqnarray}
L(\tau)
&=&
\frac{1}{2}(s+1)^{-1}
\sum_{n=0}^{s}
\sum_{m=0}^{s}
\sum_{\substack{{m'=0,}\\{m-m'=\mbox{\scriptsize odd}}}}^{s}
\tilde{\alpha}(n)
\tilde{\alpha}^{*}(m)
(-2)
\frac{\exp[i(m-m')\phi_{0}]}{\exp[i(m-m')\pi/(N+1)]-1} \nonumber \\
&&
\times
\exp[-\Gamma \tau(n+m')-i\omega(n-m')\tau]
\sum_{l=0}^{\mbox{\scriptsize min}(n,m')}
\Bigl[
\exp[2\Gamma t]-1
\Bigr]^{l}
\sqrt{
\left(
\begin{array}{c}
n \\
l \\
\end{array}
\right)
\left(
\begin{array}{c}
m' \\
l \\
\end{array}
\right)
}
|n-l\rangle\langle m'-l|.
\label{L-tau}
\end{eqnarray}
Thus, we can describe $\mbox{Tr}[\Pi^{(\pm)}_{s}L(\tau)]$ explicitly in the form,
\begin{eqnarray}
\mbox{Tr}[\Pi^{(\pm)}_{s}L(\tau)]
&=&
\frac{1}{4}(s+1)^{-1}
\sum_{m=0}^{s}
\sum_{m'=0, m-m'=\mbox{\scriptsize odd}}^{s}
\tilde{\alpha}(m')
\tilde{\alpha}^{*}(m)
(-2)
\frac{\exp[i(m-m')\phi_{0}]}{\exp[i(m-m')\pi/(N+1)]-1} \nonumber \\
&&
\pm
\frac{1}{2}(s+1)^{-2}
\sum_{m=0}^{s}
\sum_{m'=0, m-m'=\mbox{\scriptsize odd}}^{s}
\sum_{n=0, n-m'=\mbox{\scriptsize odd}}^{s}
(-2)
\frac{\exp[-i(n-m')\phi_{0}]}{\exp[-i(n-m')\pi/(N+1)]-1} \nonumber \\
&&
\times
\tilde{\alpha}(n)
\tilde{\alpha}^{*}(m)
(-2)
\frac{\exp[i(m-m')\phi_{0}]}{\exp[i(m-m')\pi/(N+1)]-1} \nonumber \\
&&
\times
\exp[-\Gamma \tau(n+m')-i\omega(n-m')\tau]
\sum_{l=0}^{\mbox{\scriptsize min}(n,m')}
[\exp(2\Gamma t)-1]^{l}
\sqrt{
\left(
\begin{array}{c}
n \\
l \\
\end{array}
\right)
\left(
\begin{array}{c}
m' \\
l \\
\end{array}
\right)
}.
\label{TrPplusminusLtau-0}
\end{eqnarray}
Based on the above calculations,
we obtain Eq.~(\ref{Trpm-L-tau}).

\end{document}